\documentclass[fleqn,usenatbib]{mnras}

\usepackage{newtxtext,newtxmath}

\usepackage[T1]{fontenc}
\usepackage{ae,aecompl}

\usepackage{graphicx}	
\usepackage{amsmath}	
\usepackage{amssymb}	
\usepackage{nicefrac}   
\usepackage{gensymb}
\usepackage{threeparttable, soul}
\graphicspath{ {figures/} }

\newcommand{\rsh}{{r_{\rm sh}}}

\newcommand{\wbv}{\omega_{\rm BV}}
\newcommand{\tadv}{t_{\rm adv}}
\newcommand{\tconv}{t_{\rm conv}}

\newcommand{\Ma}{\mathrm{Ma}}

\newcommand{\etaconvz}{\eta_{\mathrm{conv,} z}}
\newcommand{\etaconvxy}{\eta_{\mathrm{conv,} xy}}
\newcommand{\baretaconvz}{\bar{\eta}_{\mathrm{conv,} z}}
\newcommand{\baretaconvxy}{\bar{\eta}_{\mathrm{conv,} xy}}
\newcommand{\Exy}{E_{\mathrm{kin,}xy}}
\newcommand{\Ex}{E_{\mathrm{kin,}x}}
\newcommand{\Ez}{E_{\mathrm{kin,}z}}

\title[Perturbations and neutrino-driven convection]{The impact of progenitor asymmetries on the neutrino-driven convection in core-collapse supernovae}

\author[Kazeroni \& Abdikamalov]{
R\'emi Kazeroni$^{1}$\thanks{E-mail: kazeroni@MPA-Garching.MPG.DE}
and Ernazar Abdikamalov$^{2}$\thanks{E-mail: ernazar.abdikamalov@nu.edu.kz}
\\
$^{1}$Max-Planck-Institut f\"ur Astrophysik, Karl-Schwarzschild-Str. 1, D-85748 Garching, Germany \\
$^{2}$Department of Physics, Nazarbayev University, Nur-Sultan 010000, Kazakhstan
}

\date{Accepted 2020 March 30. Received 2020 March 23; in original form 2019 November 19}

\pubyear{2019}

\begin{document}
\label{firstpage}
\pagerange{\pageref{firstpage}--\pageref{lastpage}}
\maketitle

\begin{abstract}
The explosion of massive stars in core-collapse supernovae may be aided by the convective instabilities that develop in their innermost nuclear burning shells. The resulting fluctuations support the explosion by generating additional turbulence behind the supernova shock. It was suggested that the buoyant density perturbations arising from the interaction of the pre-collapse asymmetries with the shock may be the primary contributor to the enhancement of the neutrino-driven turbulent convection in the post-shock region. Employing three-dimensional numerical simulations of a toy model, we investigate the impact of such density perturbations on the post-shock turbulence. We consider a wide range of perturbation parameters. The spatial scale and the amplitude of the perturbations are found to be of comparable importance. The turbulence is particularly enhanced when the perturbation frequency is close to that of the convective turnovers in the gain region. Our analysis confirms that the buoyant density perturbations is indeed the main source of the additional turbulence in the gain region, validating the previous order-of-magnitude estimates.
\end{abstract}




\section{Introduction}
\label{sec:intro}

Multidimensional hydrodynamical effects are crucial to power the explosion of massive stars, known as core-collapse supernovae (CCSNe). The gravitational contraction of the iron core of the progenitor into a proto-neutron star triggers a bounce of the core which produces a stalled shock wave at a radius of $\sim$ 150 km. A successful explosion depends on the revival of this shock by neutrino heating during the first second after the core bounce (see \citealt{foglizzo15, janka16, mueller16b} for recent reviews). Except for the lightest progenitors, explosions of massive stars fail in spherical symmetry \citep{liebendoerfer01, kitaura06}. The shock revival relies on the decisive action of multidimensional fluid instabilities in the post-shock region, such as neutrino-driven convection \citep{herant92, herant94, burrows95, janka96, foglizzo06} and the Standing Accretion Shock Instability (SASI) \citep{blondin03, foglizzo07, foglizzo09, guilet12}. These instabilities generate large-scale turbulent motions which enhance the efficiency of the neutrino heating and ultimately trigger an asymmetric explosion. 

The neutrino mechanism is supported by an increasing number of successful three-dimensional (3D) CCSN simulations across a wide range of progenitor masses \citep{takiwaki12, takiwaki14, lentz15, melson15a, melson15b, roberts16, mueller17, mueller19, ott18, burrows19, glas19, vartanyan19}. However, it is not yet clear if the 3D explosions are powerful enough to reproduce observables such as the explosion energy or the distribution of neutron star masses \citep[e.g.][]{mueller16b}. Various ingredients leading to more robust explosions have been considered, including stellar rotation \citep{takiwaki16, kazeroni17, summa18}, modifications in the opacities \citep{melson15b}, the inclusion of muons \citep{bollig17} or finer numerical resolution \citep{nagakura19, melson20}.

Another ingredient comes from the asymmetries in the progenitors which could arise naturally in the innermost burning shells of massive stars prior to the collapse of the iron core. Linear theories suggest that fluctuations can be amplified during the collapse \citep{kovalenko98, lai00, takahashi14}. The accretion of these pre-collapse fluctuations generates additional turbulent pressure in the post-shock medium which provides better conditions for shock revival \citep{couch13b, couch15}. Aspherical accretion could lead to strong shock deformations which may enhance post-shock turbulence even further \citep{mueller15a}. Pre-collapse perturbations may reduce the critical neutrino luminosity required for an explosion by up to $\sim 20\%$ \citep{mueller16a, mueller17, radice18}. Through an extensive set of axisymmetric simulations, \citet{mueller15a} showed that the pre-collapse convective motions dominated by large spatial scales are particularly helpful to revive the stalled shock. Such conditions may be encountered in the O-burning shell and to some degree in the Si-burning shell for a large range of supernova progenitors \citep{collins18}. 3D simulations of pre-collapse shell convection revealed a wide variety of the spatial scales and strengths of these fluctuations \citep{couch15b, mueller16a, mueller19, yoshida19, yadav20}. 

Depending on their parameters, the impact of the progenitor perturbations on shock revival ranges from limited to significant. \citet{couch15b} obtained velocity fluctuations dominated by modes $\ell=\mbox{4--5}$ in a simulation of Si-burning performed in octant symmetry. The 3D initial conditions led to a slightly earlier shock revival in the subsequent CCSN simulation with a minor impact on the heating conditions. \citet{mueller16a} obtained a dominant mode $\ell=2$ in a simulation of O-burning whose impact was decisive to revive the shock in the subsequent CCSN, since the spherically symmetric progenitor failed at producing an explosion \citep{mueller17}. 

The physics of the coupling between pre-collapse perturbations and post-shock instabilities still remains to be fully understood. Using linear analysis, \citet{takahashi16} showed that the accreting perturbations may resonantly amplify SASI (or neutrino-driven convection) only if the perturbation frequency matches the oscillation frequency of SASI (or the convective growth rate). However, such a fine matching is unlikely to be common in nature. As the pre-collapse convective perturbations descend towards the centre of the star, they generate pressure waves \citep{mueller15a, abdikamalov20}. The interaction of these perturbations with the supernova shock creates acoustic, entropy, and vorticity waves in the post-shock region \citep{abdikamalov16, abdikamalov18, huete18}. Of these three, initially the vorticity waves lead to the largest increase of the non-radial motion in the post-shock region. However, the entropy waves, which are accompanied by density variations (in order to maintain pressure equilibrium with their surrounding), become buoyant and generate additional turbulence \citep{mueller16a,mueller17}. \cite{mueller16a} estimated that this is the dominant source of additional turbulence in the gain region. 

In this study, we investigate the impact of the buoyant entropic density waves on neutrino-driven convection in the gain region. We perform numerical simulations of an idealized model in which parametrized density perturbations are injected into a stationary flow that is unstable to neutrino-driven convection \citep[][hereafter Paper I]{kazeroni18}. Since entropic density waves are generated in the post-shock region after the interaction of the pre-collapse upstream perturbations with the shock, we do not include the shock wave in our idealized model. This simplification allows us to disentangle the impact of the entropic density perturbations on the neutrino-driven convection from the interaction of the upstream perturbations with the shock. Controlled simulations enable us to quantify the impact of the perturbation properties, such as the amplitude, the spatial scale, and the temporal frequency on turbulence. Finally, we perform additional two-dimensional (2D) simulations and examine their limitations by comparing them to our 3D models. For brevity, hereafter we will refer to entropic density waves as density perturbations.

Our results confirm that the injection of turbulent kinetic energy in the gain layer mostly results from the work of buoyancy on the density perturbations. The amount of turbulent kinetic energy generated by the perturbation can well be captured by an analytical estimate. The spatial scale and the amplitude of the perturbation have a comparable impact on the additional turbulent kinetic energy. We find that turbulence can be significantly enhanced if the perturbation frequency is close to that of convective turnovers. While large-scale modes are found to maximize the turbulent kinetic energy both in 2D and 3D, the strength of the turbulence is largely overpredicted in 2D due to a different turbulent energy cascade.

The structure of the paper is the following. The methodology is detailed in Section \ref{sec:model}. The results of the numerical simulations are described in Section \ref{sec:results}. The origin and the properties of the additional turbulence related to the accretion of density perturbations are investigated in Section \ref{sec:analysis}. The discrepancies between 2D and 3D simulations are analysed in Section \ref{sec:dimension}. We summarize and discuss our findings in Section \ref{sec:conclusion}.

\section{Physical and numerical setup}
\label{sec:model}

\subsection{Stationary flow}
\label{subsec:flow}

As in Paper I, we consider a stationary flow of an ideal gas with ${\gamma=4/3}$ along vertical $z$-direction in a 3D Cartesian geometry. The computational domain covers the region ${-150\,\mathrm{km} \leq x,y \leq 150\,\mathrm{km}}$ and ${-450\,\mathrm{km} \leq z \leq 450\,\mathrm{km}}$. The central ${-H \leq z \leq H}$ region, where ${H=50\,\rm{km}}$, mimics the gain layer of a CCSN, where neutrino absorption exceeds neutrino emission (cf. Fig.~\ref{fig:pert_sketch}). The heating and the gravity in this layer are modelled as described Paper I. The heating is proportional to density and is normalized by the coefficient ${K_{\rm H}=2.76\times10^{-3}}$, while the gravitational potential $\nabla \Phi$ is normalized by the coefficient ${K_{\rm G}=3}$ (cf. Paper I for the definitions of these quantities). Outside the gain layer, heating and gravity vanish and the stationary flow is uniform. The vertical boundaries are placed far enough to minimize the impact of reflected acoustic waves. The upstream Mach number is set to ${\Ma_{\rm up} = 0.3}$, where the subscript ``up'' refers to values in the upstream flow. As mentioned earlier, in order to study the coupling between neutrino-driven convection and injected perturbations in its simplest form, our model does not contain a shock wave to avoid any feedback that would result from its deformation. In addition, the cooling layer surrounding the proto-neutron star is absent from our model for the sake of simplicity.

For convection to develop in the gain layer, the buoyancy has to operate faster than advection through the gain layer. This condition can be expressed in terms of the parameter ${\chi}$ \citep{foglizzo06}, which measures the ratio of the advection time-scale through the gain region $\tadv$ to the convective turnover time-scale $\tconv$:
\begin{equation}
 \label{eq:chi}
 \chi \equiv \int_{-H}^{H}\left| \mathrm{Im}\left(\langle\wbv^2\rangle^{1/2}\right) \dfrac{\mathrm{d}z}{\langle v_z\rangle}\right|,
\end{equation}
where $\langle \cdot \rangle$ refers to horizontally averaged quantities and $v_z$ to the vertical velocity. The advection time-scale through the gain layer is
\begin{equation}
 \label{eq:tadv}
 \tadv \equiv \int_{-H}^H \dfrac{dz}{\left| \langle v_z \rangle \right|}.
\end{equation}
while convective turnover time-scale can be approximated as ${\tconv\sim\wbv^{-1}}$, where the Brunt-V\"ais\"al\"a frequency $\wbv$ is defined as
\begin{equation}
 \label{eq:wbv}
 \wbv \equiv \left(\nabla \Phi \right)^{1/2} \left|\dfrac{\nabla P}{\gamma P} - \dfrac{\nabla \rho}{\rho} \right|^{1/2} = \left( \dfrac{\gamma-1}{\gamma}\nabla \Phi \nabla S\right)^{1/2},
\end{equation}
with $P$, $\rho$, and $S$ being respectively the pressure, the density, and the dimensionless entropy of the flow. The latter is computed as
\begin{equation}
 \label{eq:Sdef}
 S \equiv \dfrac{1}{\gamma-1} \log{\left[ \left(\dfrac{P}{P_{\rm up}}\right) \left(\dfrac{\rho_{\rm up}}{\rho}\right)^{\gamma}\right]}.
\end{equation}

In our setup, the stationary flow is linearly unstable to neutrino-driven convection only if ${\chi \gtrsim \chi_{\rm crit}\approx 2.4}$ \citep{kazeroni18}\footnote{Note that for a more realistic model including a shock wave, the instability threshold is ${\chi_{\rm crit} \approx 3.2}$ \citep{foglizzo06}.}. For our choice of parameters $K_{\rm G}$, $K_{\rm H}$, and $\Ma_{\rm up}$, the stationary flow has ${\chi=3}$, i.e. ${\tconv \approx \tadv/3}$, which is slightly above the instability threshold.  

\subsection{Perturbation modelling}
\label{subsec:perturbation}

\begin{figure}
\centering
	\includegraphics[width=0.78\columnwidth]{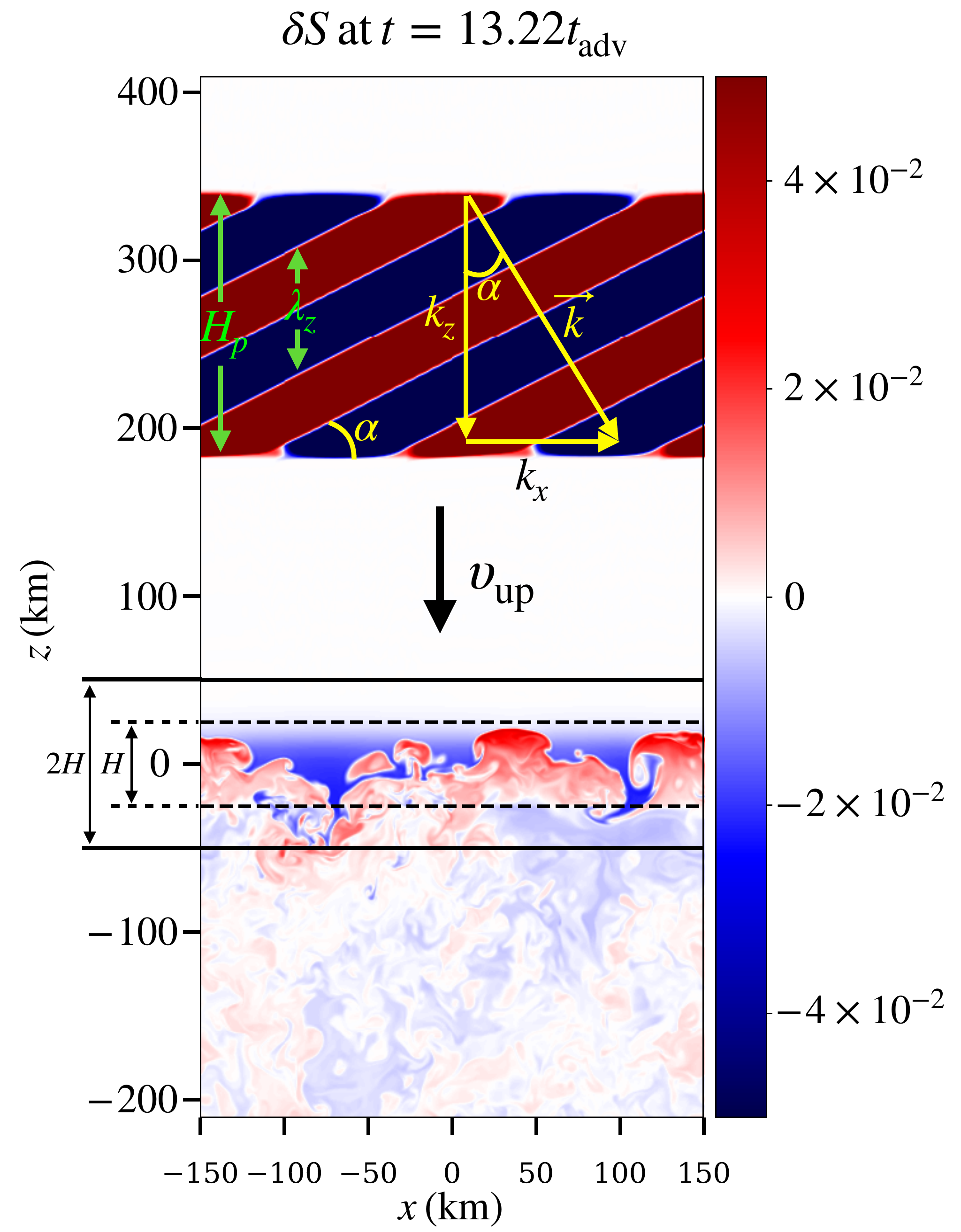}
    \caption{Structure of the perturbation, in the plane ${y=0\,\rm{km}}$, shown in entropy contrast $\delta S = S-\langle S\rangle$. At a time ${t=13.22\,\tadv}$, the neutrino-driven convection is fully turbulent in the gain layer located between ${-H \leq z \leq H}$. Outside the two horizontal solid lines, gravity and heating are turned off, while they are at their full intensity inside the horizontal dashed lines. The flow above the gain layer has a uniform velocity $v_{\rm up}$. The injected perturbation is characterized by a height $H_{\rm p}$, a vertical spatial wavelength $\lambda_z$, and horizontal and vertical wavenumbers respectively labelled $k_x$ and $k_z$. The angle $\alpha $ corresponds to the tilt angle from the horizontal axis to the direction of the entropy bands.
    }
    \label{fig:pert_sketch}
\end{figure}

We first generate a flow where neutrino-driven convection has reached a fully turbulent stage. A simulation of neutrino-driven convection is performed starting from a vertical stationary flow and using a density perturbation of $0.1\%$ amplitude. The non-linear stage of neutrino-driven convection is reached after 10 advection times. The simulation is run for a total duration of 18 advection times. This simulation is used as a reference model to investigate the impact of injected perturbations on the convective instability. 

We then take an output of the reference simulation at a time ${t=12.8\,\tadv}$ and inject a parametrized density perturbation through the upper boundary of domain, located at ${z=450\,\rm{km}}$. At this time, neutrino-driven convection in the gain layer has reached the non-linear regime for about than 9 convective turnover times. The simulations of perturbed convection are evolved from ${t=12.8\,\tadv}$ to ${t=18\,\tadv}$ which covers about 15 convective turnover times. We find that the simulations reach a steady state regime by this time and a further evolution does not bring any different outcome. 

The infalling density perturbation is modelled as a planar sinusoidal structure in the $xz$-plane (Fig.~\ref{fig:pert_sketch}):
\begin{equation}
\label{eq:drho}
    \dfrac{\delta\rho}{\rho_{\rm up}} = A\sin{\left[k_x x + k_z z - \omega t\right]},
\end{equation}
where $\rho_{\rm up}$ is the the density of the upstream stationary flow. In order to maintain the pressure equilibrium with the surrounding flow, the associated temperature variations have the opposite phase. The parameter $A$ is the amplitude, $k_x$ and $k_z$ are the horizontal and vertical components of the wavenumber vector $\vec{k}$, and $\omega$ is the temporal frequency. 
The horizontal wavenumber is computed as
\begin{equation}
\label{eq:kx}
    k_x = \dfrac{2\pi n_x}{L_x},
\end{equation}
where ${L_x=300\,\rm{km}}$ is the horizontal extent of the gain region.  The integer $n_x$ corresponds to the number of wavelengths in the $x$-direction, which plays a role similar to the angular wavenumber $\ell$ of the spherical harmonics representation. 
The vertical wavenumber is given by
\begin{equation}
 \label{eq:kz}
    k_z = k_x \cot{\left(\alpha\right)}, 
\end{equation}
where $\alpha$ corresponds to the tilt angle from the $x$-axis to the direction of the entropy/density bands as shown in Fig.~\ref{fig:pert_sketch}. The vertical spatial wavelength can be obtained by
\begin{equation}
 \label{eq:lambda_z}
 \lambda_z = \dfrac{2\pi}{k_z} = \dfrac{L_x}{n_x}\tan{\left(\alpha\right)}.
\end{equation}
Besides, the temporal frequency $\omega$ is given by
\begin{equation}
\label{eq:omega}
    \omega = k_z |v_{\rm up}| = k_x \cot{\left(\alpha\right)} |v_{\rm up}| = ||\vec{k}|| \cos{\left(\alpha\right)} |v_{\rm up}|,
\end{equation}
with $v_{\rm up}$ being the vertical velocity of the upstream uniform flow. 
The density perturbation can thus be rewritten as\footnote{Note that in the limit ${\alpha=0\degree}$, the density perturbation can be expressed as ${\delta\rho/\rho_{\rm up}=A\sin{\left[k_z\left(z-|v_{\rm up}|t\right)\right]}}$.} 
\begin{equation}
    \dfrac{\delta\rho}{\rho_{\rm up}} = A\sin{\left[\frac{2\pi n_x}{L_x}\left[x+\cot{\left(\alpha\right)}\left(z-|v_{\rm up}|t\right)\right]\right]}.
\end{equation}

The perturbation is injected through the upper boundary for a finite duration of time defined as 
\begin{equation}
 \label{eq:tinj}
 \Delta T_{\rm inj} = \dfrac{H_{\rm p}}{|v_{\rm up}|},
\end{equation}
where $H_{\rm p}$ is the vertical extent of the perturbation (Fig.~\ref{fig:pert_sketch}). In the following, $H_{\rm p}$ is normalized by the vertical spatial wavelength $\lambda_z$.

In our study, the free parameters of the injected perturbations are chosen to be $A$, $n_x$, $\alpha$, and $H_{\rm p}$. In the rest of the paper, we will consider the following parameter space: ${5\% \leq A \leq 20\%}$, ${30\degree \leq \alpha \leq 90\degree}$, ${1 \leq n_x \leq 8}$, and ${0.25 \leq H_{\rm p}/\lambda_z \leq 4}$, or correspondingly ${1/32 \leq \Delta T_{\rm inj}/\tadv \leq 1}$. This parameter space covers a wide range of perturbation properties, including those that where observed in numerical simulations \citep{couch15b,mueller16a}. This will allow us to carefully study the impact of each of these parameters on the evolution of the system. 

\subsection{Numerical simulations}
\label{subsec:simulations}

  \begin{table}
  \caption{Overview of the simulations discussed in this work.}
  \label{tab:simulations}
  \begin{threeparttable}
  \begin{tabular}{lccccccc}
  \hline
  Model \tnote{a} & $A$ \tnote{b} & $n_x$ \tnote{c} & $\alpha$ \tnote{d} & $H_{\rm p}$ \tnote{e} & $\Delta\, T_{\rm inj}$ \tnote{f} & $\lambda_z$ \tnote{g} & $\omega$ \tnote{h} \\
  & $(\%)$ &  & $(\degree)$ & $(\lambda_z)$ & $(\tadv)$ & $(H)$ & $(\omega_{\rm conv})$\\
  \hline
  \hline
  		Hp0.25 & 10 & 2 & 30 & 0.25 & 1/32 & 0.8 & 2.29 \\
		Hp0.5 & 10 & 2 & 30 & 0.5 & 1/16 & 0.8 & 2.29 \\
		Hp1 & 10 & 2 & 30 & 1 & 1/8 & 0.8 & 2.29 \\
		Hp2 & 10 & 2 & 30 & 2 & 1/4 & 0.8 & 2.29 \\
		Hp4 & 10 & 2 & 30 & 4 & 1/2 & 0.8 & 2.29 \\
		Hp8 & 10 & 2 & 30 & 8 & 1 & 0.8 & 2.29 \\
    \hline
        A5$\_$nx1 & 5 & 1 & 30 & 2 & 0.5 & 1.6 & 2.29 \\
		A5$\_$nx2 & 5 & 2 & 30 & 4 & 0.5 & 0.8 & 2.29 \\
		A5$\_$nx4 & 5 & 4 & 30 & 8 & 0.5 & 0.4 & 2.29 \\
	    A7.5$\_$nx3 & 7.5 & 3 & 30 & 6 & 0.5 & 0.53 & 2.29 \\
	    A7.5$\_$nx6 & 7.5 & 6 & 30 & 12 & 0.5 & 0.27 & 2.29 \\
		A10$\_$nx1 & 10 & 1 & 30 & 2 & 0.5 & 1.6 & 2.29 \\
		A10$\_$nx2 & 10 & 2 & 30 & 4 & 0.5 & 0.8 & 2.29 \\
		A10$\_$nx4 & 10 & 4 & 30 & 8 & 0.5 & 0.4 & 2.29 \\
		A10$\_$nx8 & 10 & 8 & 30 & 20 & 0.5 & 0.2 & 2.29 \\
	    A12.5$\_$nx5 & 12.5 & 5 & 30 & 10 & 0.5 & 0.32 & 2.29 \\
	    A15$\_$nx3 & 15 & 3 & 30 & 6 & 0.5 & 0.53 & 2.29 \\
	    A15$\_$nx6 & 15 & 6 & 30 & 12 & 0.5 & 0.27 & 2.29 \\
		A20$\_$nx2 & 20 & 2 & 30 & 4 & 0.5 & 0.8 & 2.29 \\
		A20$\_$nx4 & 20 & 4 & 30 & 8 & 0.5 & 0.4 & 2.29 \\
	\hline
        ang30 & 10 & 2 & 30 & 4 & 0.5 & 0.8 & 2.29\\
		ang45 & 10 & 2 & 45 & 2.1 & 0.5 & 1.5 & 1.32 \\
		ang60 & 10 & 2 & 60 & 1.2 & 0.5 & 2.6 & 0.76\\
		ang75 & 10 & 2 & 75 & 0.6 & 0.5 & 5.6 & 0.35\\
		ang90 & 10 & 2 & 90 & - & 0.5 & - & - \\
  	\hline
  \end{tabular}
  \begin{tablenotes}
    \item[a] Simulation name.
    \item[b] Perturbation amplitude.
    \item[c] Horizontal wavenumber.
    \item[d] Tilt angle.
    \item[e] Perturbation height normalized by the vertical spatial wavelength (Eq.~\ref{eq:lambda_z}).
    \item[f] Injection time-scale normalized by the advection time (Eq.~\ref{eq:tinj}).
    \item[g] Vertical spatial wavelength normalized by the gain region height $H$.
    \item[h] Temporal frequency of the perturbation normalized by the convective turnover frequency (Eqs.~\ref{eq:omega} and~\ref{eq:omega_conv}).
  \end{tablenotes}

  \end{threeparttable}
 \end{table}

The numerical setup employed in this study is similar to the one of Paper I. The RAMSES code \citep{teyssier02, fromang06} is used to simulate the flow dynamics on a Cartesian grid. At the top and bottom boundaries, we use a constant inflow and outflow conditions set by the stationary flow. We use periodic boundary conditions at the lateral edges. The domain is respectively decomposed into ${(N_x \times N_z) = (400 \times 1200)}$ and ${(N_x \times N_y \times N_z) = (400 \times 400 \times 1200)}$ uniform cells in 2D and in 3D simulations. Except in Section~\ref{sec:dimension}, which focuses on the impact of dimensionality, all of our simulations are performed in 3D.

In Table~\ref{tab:simulations}, we present the full list of the simulations discussed in this work. We evolve three distinct sequences of models where we vary one or two perturbation parameters at a time. The first set of models is used to examine the influence of the perturbation height and the simulations are named according to the following convention. In the models labelled HpX, the height of the perturbation is set to ${H_{\rm p}=X\lambda_z}$ with X ranging from 0.25 to 8. The corresponding values of  injection time-scales range from $1/32\,\tadv$ to $\tadv$. The other parameters are set to ${A=10\%}$, ${n_x=2}$, and ${\alpha=30\degree}$.

Since the impact of the perturbation horizontal wavenumber and amplitude are expected to be of the same order \citep{mueller15a, mueller17}, in our second sequence of models we vary both $A$ and $n_x$. These simulations are named AX$\_$nxY, where X corresponds to the perturbation amplitude, ranging from $5\%$ to $20\%$, and Y to the horizontal wavenumber, ranging from 1 to 8. The tilt angle is set to ${\alpha = 30\degree}$ and the perturbation height to ${H_{\rm p} = 4\lambda_z}$.

Finally, the impact of the tilt angle $\alpha$ on the flow dynamics is investigated using the simulations labelled as angX where X denotes the value of the angle, ranging from $30\degree$ to $90\degree$. In these simulations, the remaining parameters are set to ${A=10\%}$, ${n_x=2}$, and ${H_{\rm p} = 4\lambda_z}$.

\section{Overview of the simulation results}
\label{sec:results}

\subsection{Non-linear neutrino-driven convection}
\label{subsec:convection}

\begin{figure}
\centering
	\includegraphics[width=0.88\columnwidth]{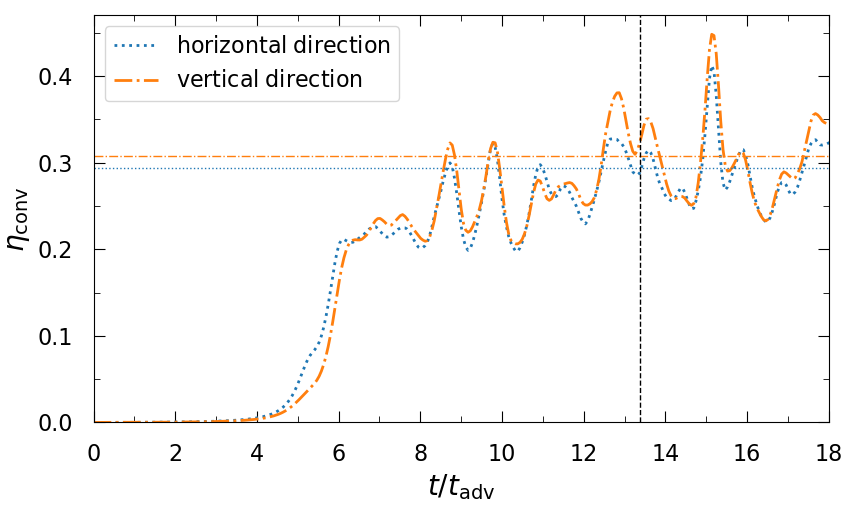}
    \caption{Horizontal (dotted curve) and vertical (dash-dotted curve) efficiency factors computed in the reference simulation as a function of time (Eqs.~\ref{eq:eta_conv_xy}-~\ref{eq:eta_conv_z}). The horizontal lines depict the mean values over the non-linear phase of the instability. The vertical dashed line marks the time $t=13.2\tadv$ when accreting perturbations enter the gain layer in cases where neutrino-driven convection is perturbed. For readability, the data are smoothed out over a convective turnover time-scale.
    }
    \label{fig:conv}
\end{figure}

In the absence of accreting perturbations, the kinetic energy of neutrino-driven convection results from the neutrino heating. The efficiency factors for the conversion of the latter to the former can be expressed in terms of the mass $M_{\rm g}$, the net neutrino heating $\mathcal{L}$, and the height $2H$ of the gain layer \citep{mueller15a}:
\begin{equation}
 \label{eq:eta_conv_xy}
 \etaconvxy \equiv \dfrac{\Exy}{\left[2H\mathcal{L}\right]^{2/3}M_{\rm g}^{1/3}},
\end{equation}
and
\begin{equation}
 \label{eq:eta_conv_z}
 \etaconvz \equiv \dfrac{\Ez}{\left[2H\mathcal{L}\right]^{2/3}M_{\rm g}^{1/3}},
\end{equation}
respectively for the horizontal and vertical components of the turbulent kinetic energy.
The horizontal turbulent kinetic energy is computed as
\begin{equation}
 \label{eq:Ekin_xy}
 \Exy = \dfrac{1}{2} \int_{V_{\rm gain}} \rho\left(v_x^2 + v_y^2\right)\mathrm{d}V,
\end{equation}
with $v_x$ and $v_y$ being the velocity components in the $x$ and $y$-directions, and $V_{\rm gain}$ being the volume of the gain region.
The vertical turbulent kinetic energy is similarly computed as
\begin{equation}
 \label{eq:Ekin_z}
 \Ez = \dfrac{1}{2} \int_{V_{\rm gain}} \rho\left(v_z - \langle v_z\rangle\right)^2 \mathrm{d}V.
\end{equation}
The total heating and mass in the gain layer are barely affected by the convective instability. As a consequence, the efficiency factors are directly proportional to the components of the turbulent kinetic energy.

Our simulation of non-perturbed neutrino-driven convection shows that both efficiency factors reach asymptotic values in the non-linear stage of the instability, defined as ${10\,\tadv \leq t \leq 18\,\tadv}$ (Fig.~\ref{fig:conv}). We find that ${\etaconvxy \approx \etaconvz \approx 0.3}$, which is consistent with the values obtained in realistic 3D simulations of CCSNe \citep{mueller17}.

\subsection{Perturbation height}
\label{subsec:duration}

\begin{figure*}
\centering
	\includegraphics[width=0.9\columnwidth]{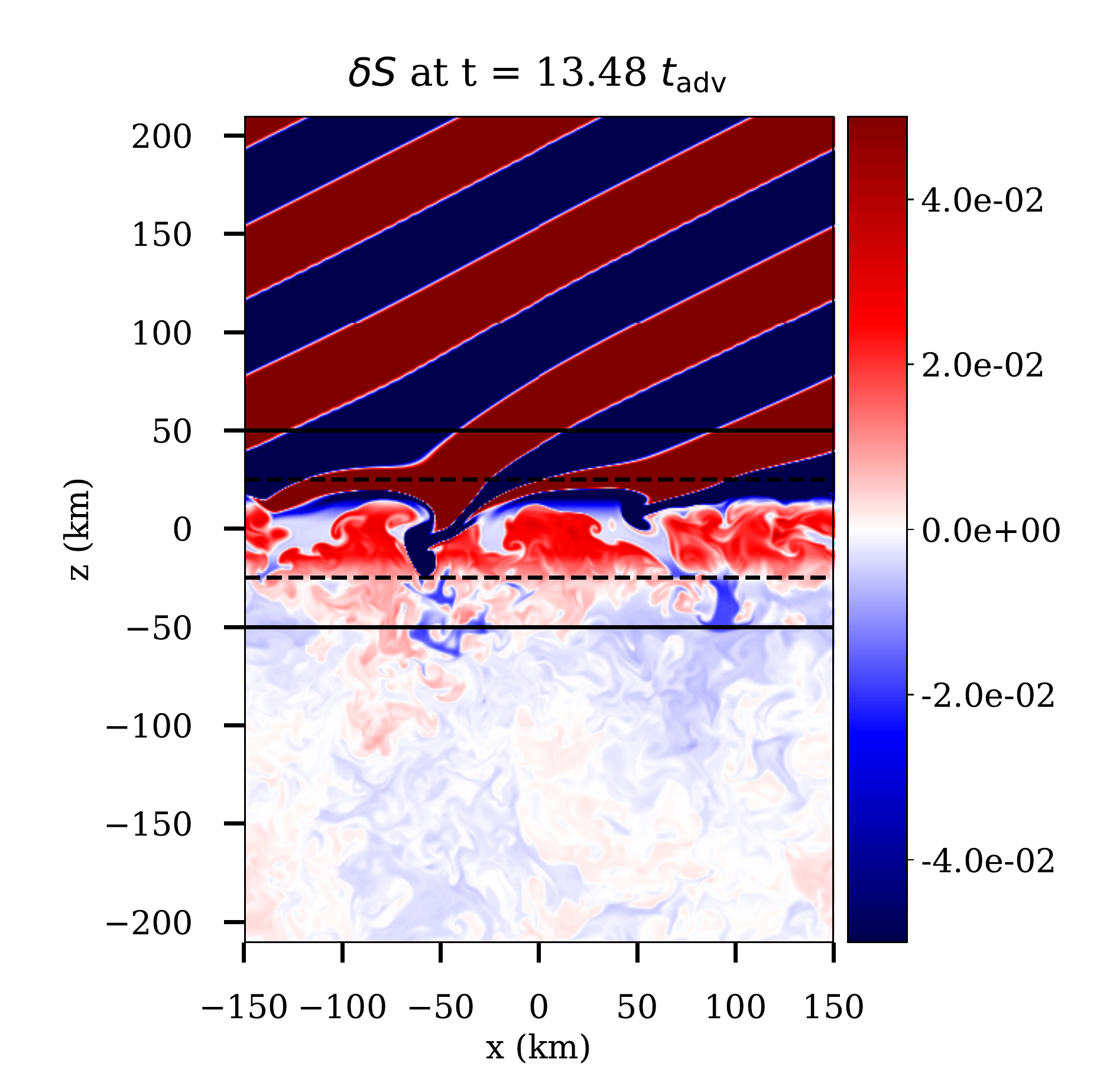}
	\includegraphics[width=0.9\columnwidth]{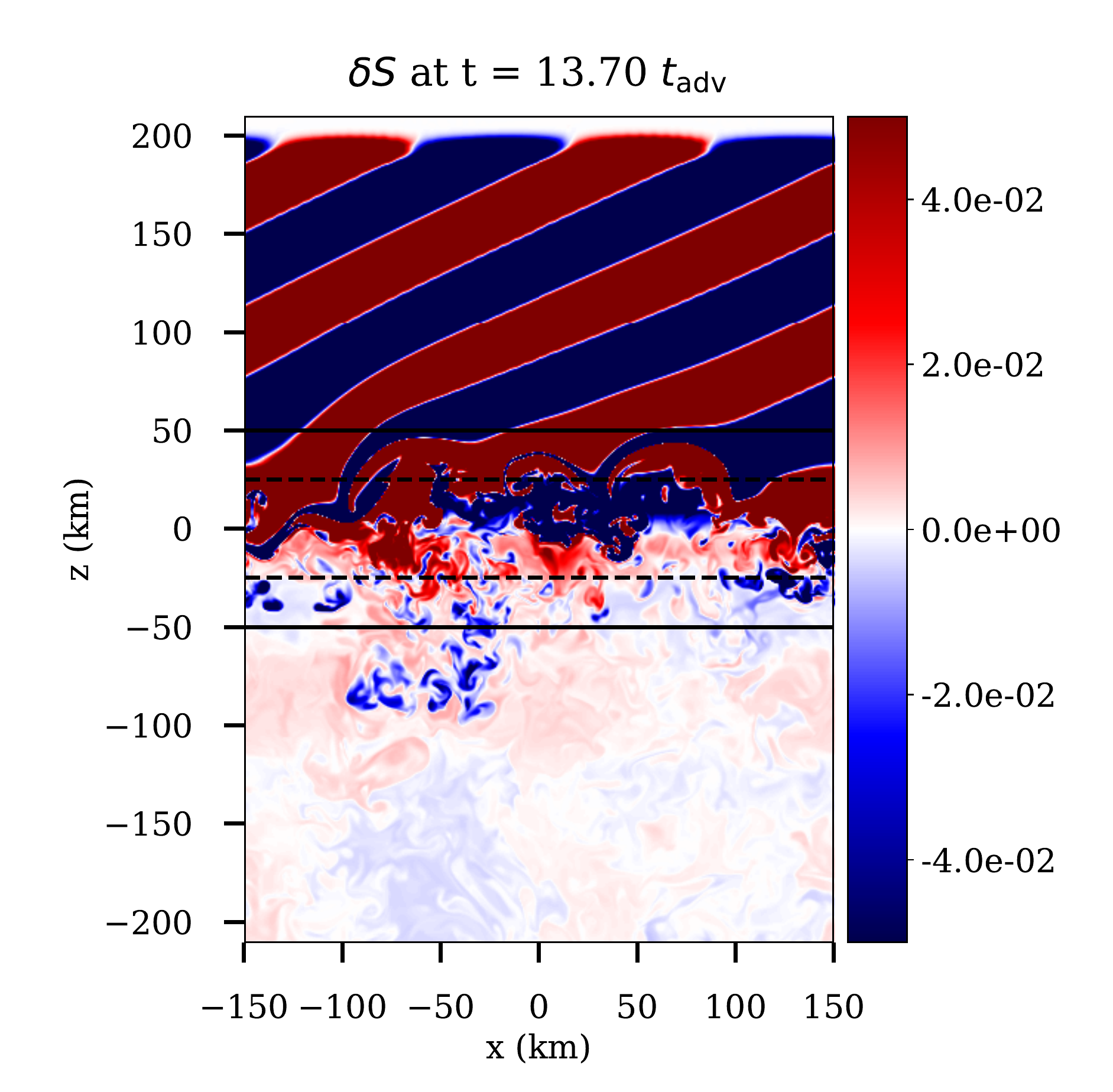}
	\includegraphics[width=0.9\columnwidth]{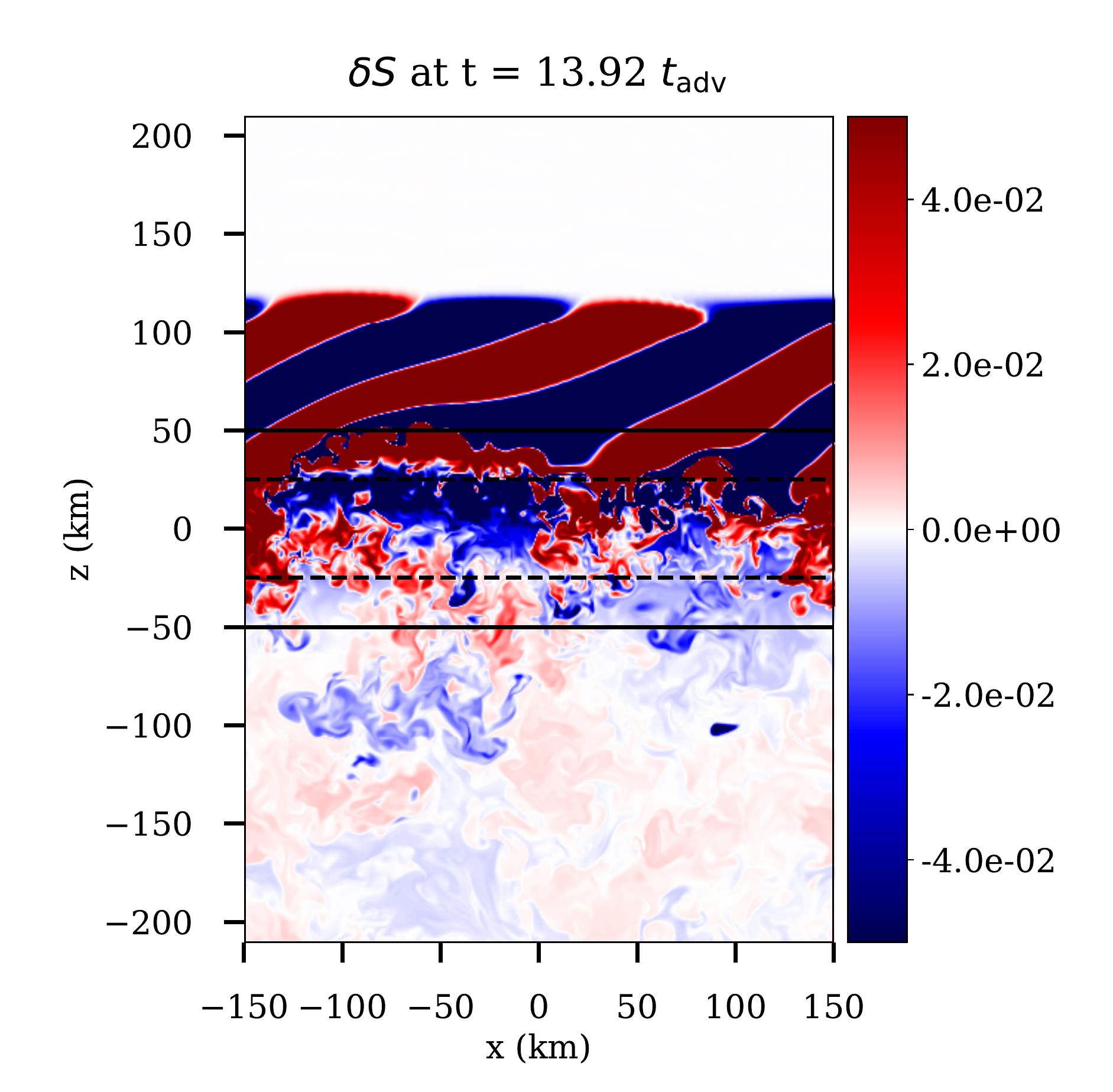}
	\includegraphics[width=0.9\columnwidth]{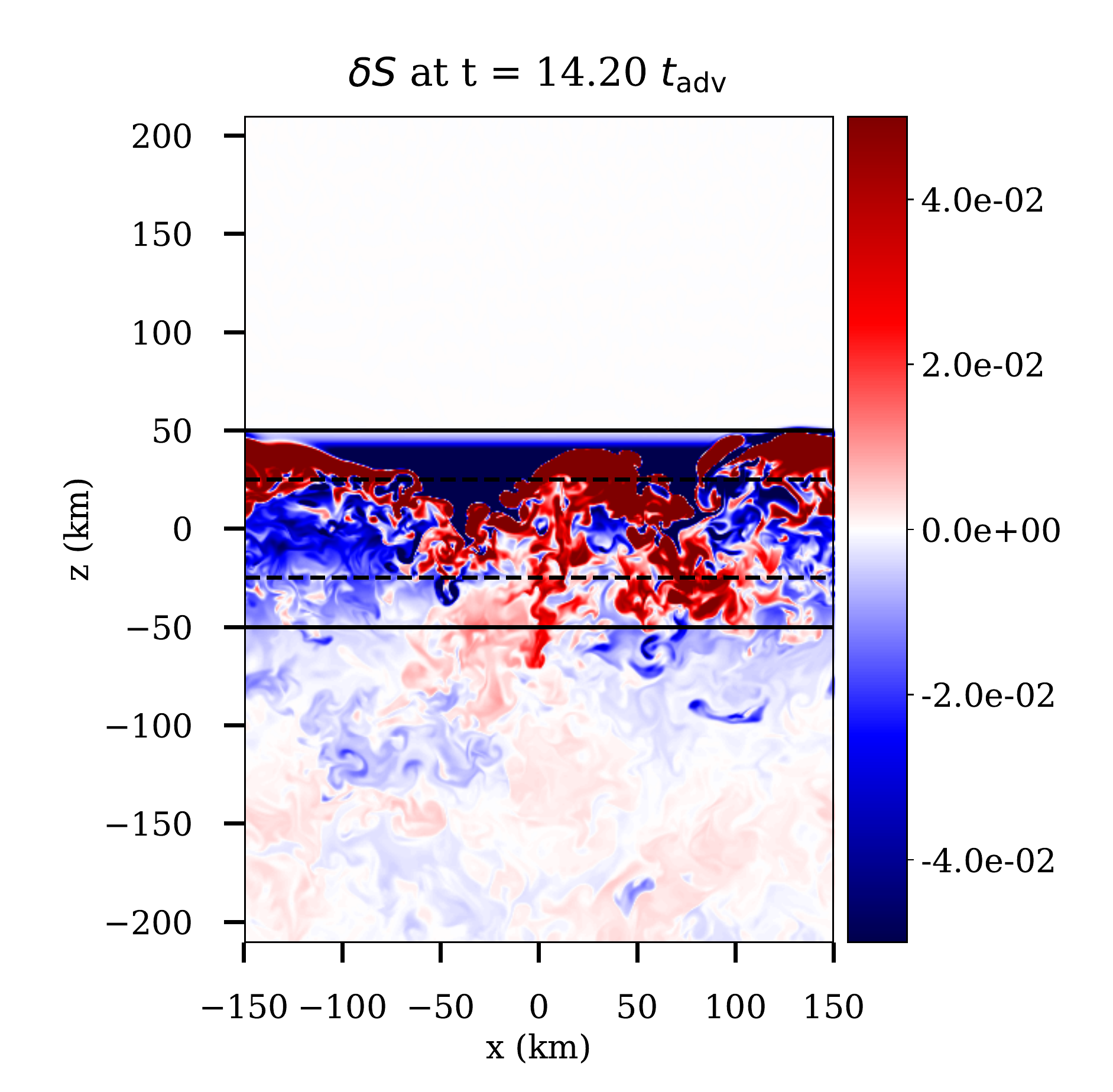}
    \caption{Snapshots of entropy contrast, ${\delta S = S -\langle S \rangle}$, shown in the plane ${y=0\, \rm{km}}$ for the model A10$\_$nx2 (see Table~\ref{tab:simulations}) after 13.48, 13.7, 13.92, and 14.2 advection time-scales (left to right, top to bottom). The horizontal lines are defined in Fig.~\ref{fig:pert_sketch}. }
    \label{fig:snapshots}
\end{figure*}

\begin{figure}
\centering
	\includegraphics[width=0.9\columnwidth]{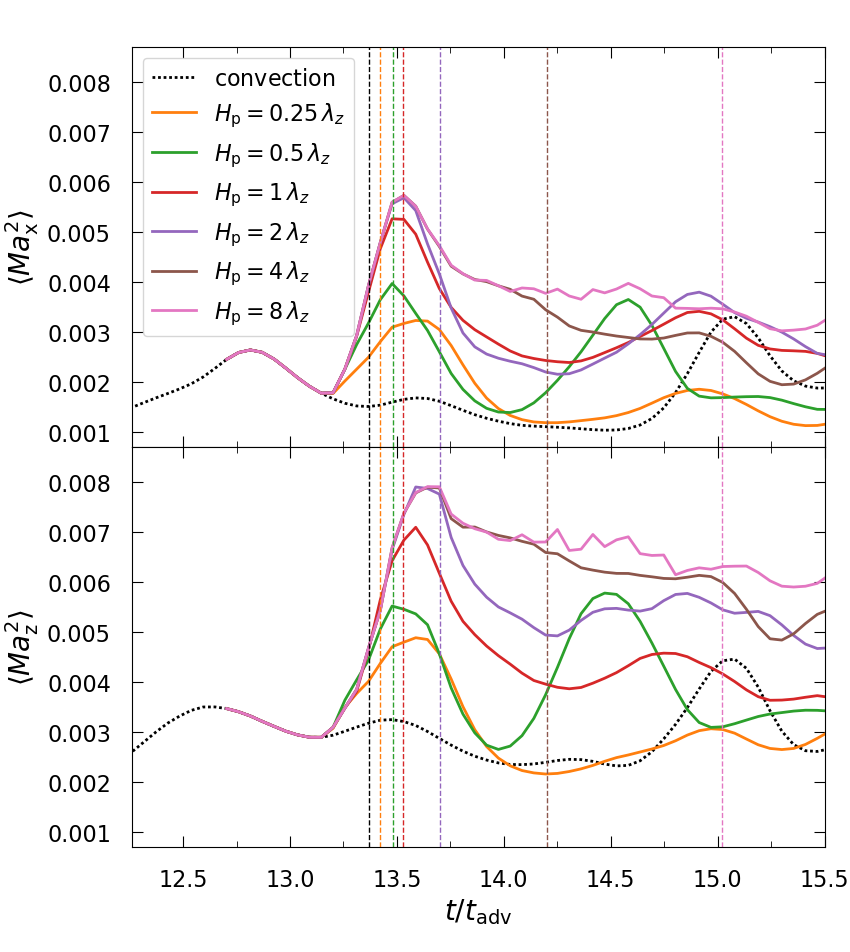}
    \caption{Horizontal (top) and vertical (bottom) squared turbulent Mach numbers shown for the non-perturbed case (black) as well as perturbed cases with different perturbation heights $H_{\rm p}$ (models HpX in Table~\ref{tab:simulations}). The leftmost vertical dashed line marks the arrival of the injected perturbation inside the gain layer (in black). The other lines depict the end of the perturbation accretion through the upper edge of the gain layer. For readability, the data are smoothed out over a convective turnover time-scale. The apparent earlier growth of the components of $\Ma^2$ is solely due to the time averaging of the quantities.}
    \label{fig:Ma2_duration}
\end{figure}

Figure~\ref{fig:snapshots} illustrates the dynamics of gain region when an injected perturbation accretes through this layer for representative model A10$\_$nx2 (see Table~\ref{tab:simulations}). As the perturbation enters the gain region, it shrinks in the vertical direction due to the deceleration of the flow. The compression factor equals the velocity reduction ratio (Eq.~\ref{eq:omega}), which amounts to ${v_{\rm up}/v_{\rm z,g}\approx 7.5}$ in our setup, where $v_{\rm z,g}$ is the average vertical velocity in the gain layer. Once an injected perturbation enters the gain layer, at ${t=13.2\,\tadv}$, it adjusts itself to the new pressure of the heating region (Top left-hand panel). The buoyant forces push the higher entropy parts of the perturbation above the layer of turbulent convection. The lateral expansion of these buoyant entropy bands generates pinched downflows of cold material which then undergo turbulent convection. As the perturbation accretion continues, the higher entropy material gradually fills the whole upper part of the gain region (Top right-hand panel). This process saturates when the first full vertical wavelength of the perturbation has entered the gain layer. The base of the perturbation is then subject to convective mixing that disrupts the entropy bands while the rest of the perturbation experiences the effects of buoyancy and pressure adjustment (Bottom left-hand panel). The accumulation of the buoyant material in the upper part of the gain layer tend to lower the altitude at which the perturbation is disrupted by turbulent convection (Bottom right-hand panel).

In order to study the impact of the perturbation height on the turbulence in the gain layer, we focus on the models with varying $H_{\rm p}$, labelled HpX in Table~\ref{tab:simulations}. We calculate the squared turbulent Mach number in the $i$-direction
\begin{equation}
 \label{eq:mach2}
 \langle \Ma^2_i \rangle = \dfrac{\langle \left( v_i - \langle v_i \rangle \right)^2 \rangle}{\langle c^2 \rangle},
\end{equation}
with $c$ being the sound speed and ${\langle v_x \rangle=\langle v_y \rangle=0}$ since the flow is initially purely vertical. Fig.~\ref{fig:Ma2_duration} shows the components of $\Ma^2$ in the $x$ and $z$-direction. The components of $\Ma^2$ begin to increase as the perturbation starts to accrete through the gain region. After a phase of identical increase in all cases, the growth becomes dependent on $H_{\rm p}$. The maximum of $\Ma^2$ increases with the perturbation height up to ${H_{\rm p} = 2 \lambda_z}$, which corresponds to an injection time ${T_{\rm inj} = \tadv/4}$. For larger values of $H_{\rm p}$, the maxima do not grow with $H_{\rm p}$ anymore. For such perturbations, the growth of $\Ma^2$ stops before the end of the perturbation accretion through the upper edge of the gain layer. This is likely a consequence of the accumulation of buoyant material in the upper part of the gain layer, as discussed above. This phenomenon tends to reduce the size of the region of neutrino-driven convection and thus to weaken turbulence. 

After reaching their maxima, the components of $\Ma^2$ remain higher than those of the non-perturbed convection for several advection times. During this post-maximum phase, we observe secondary peaks in $\Ma^2$ in some perturbed models and the non-perturbed case. These peaks are a result of the stochastic nature of the neutrino-driven convection in the gain layer. In a more realistic framework, the additional turbulent pressure generated by the accretion of extended perturbations would move the shock outwards. These perturbations may then be entirely disrupted inside the enlarged gain layer before quenching convection.

\subsection{Perturbation amplitude and horizontal wavenumber}
\label{subsec:amplitude}

\begin{figure*}
\centering
	\includegraphics[width=0.9\columnwidth]{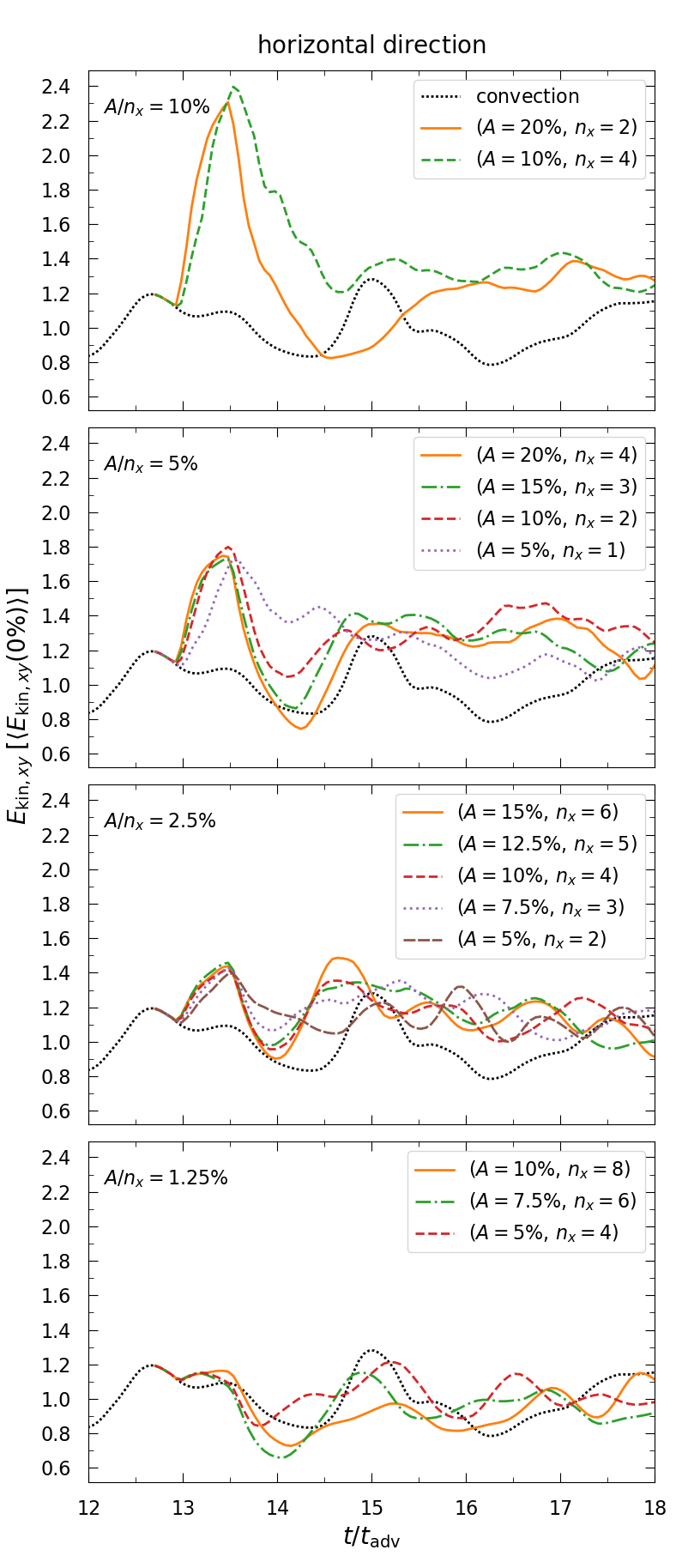}
	\includegraphics[width=0.9\columnwidth]{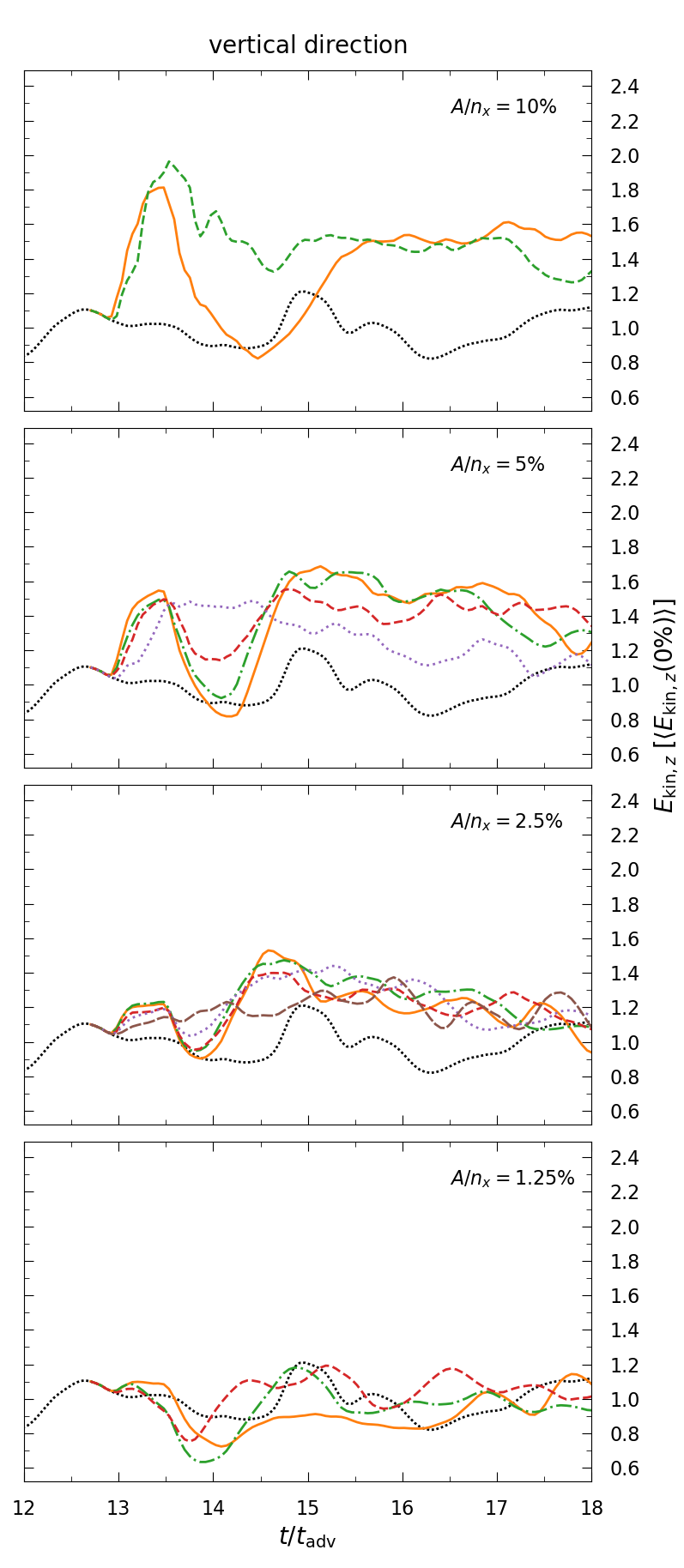}
	\caption{The horizontal (left-hand panels) and vertical (right-hand panels) components of the turbulent kinetic energy are shown for the simulations with varying $A$ and $n_x$. See the models AX$\_$nxY in Table~\ref{tab:simulations} for the rest of the parameters. 
	In each panel, the perturbed cases for a given ratio $A/n_x$ are compared to the simulation of non-perturbed neutrino-driven convection (black dotted curve). 
    All quantities are divided by the average values computed in the non-linear regime of neutrino-driven convection.
    For readability, the data are smoothed out over a turnover time-scale.}
    \label{fig:Ekin_amplitude}
\end{figure*}

Figure~\ref{fig:Ekin_amplitude} shows the impact of the perturbation amplitude $A$ and its horizontal wavenumber $n_x$ on the neutrino-driven convection. The horizontal and the vertical components of the turbulent kinetic energy in the gain layer are normalized by the averaged components measured in the reference simulation during the non-linear regime of convection, i.e. ${10 \leq t/\tadv \leq 18}$. In all perturbed models, the turbulent kinetic energy reaches an initial peak as the perturbation accretes through the gain region. The early growth of the turbulent kinetic energy is stronger for perturbations with larger amplitude because larger gravitational potential energy is converted into kinetic energy. Similarly, the turbulent kinetic energy grows faster for lower horizontal wavenumbers. This is a consequence of a larger vertical spatial wavelength over which the mixing occurs (Eq.~\ref{eq:lambda_z}).

During the accretion phase of the perturbation, doubling $A$ leads to a similar level of turbulent kinetic energy as halving $n_x$. For a given ratio $A/n_x$, all perturbed cases display very similar initial peaks of turbulent kinetic energy  (Fig.~\ref{fig:Ekin_amplitude}). This suggests that the impact of $n_x$ on the convective instability is as large as that of $A$. This implies that the level of turbulence reached in the gain layer is correlated with the ratio $A/n_x$. We defer a more detailed analysis of this conjecture to Section~\ref{sec:analysis}.

The correlation between the ratio $A/n_x$ and the components of the turbulent kinetic energy holds for both the horizontal and the vertical directions (Fig.~\ref{fig:Ekin_amplitude}). The turbulent kinetic energy is initially higher in the horizontal directions. As we will see in the next section, this behavior is specific to the particular value of the tilt angle $\alpha$ used here. When the accretion of the perturbation stops, the two components are found to be roughly equal to each other, similarly to non-perturbed convection (Paper I). 

Once the turbulent kinetic energy reaches its maximum, we observe that all components remain larger than their counterparts in the simulation of non-perturbed convection. In this post-accretion phase, the turbulent kinetic energy is larger for models with larger perturbation amplitudes, especially in the vertical direction (Fig.~\ref{fig:Ekin_amplitude}, right-hand panels). Compared to the reference model, the vertical turbulent kinetic energy is about $30\%$ and $50\%$ higher for ${A=10\%}$ and ${A=20\%}$ respectively. The parameter $n_x$ seems to play a minor role on the post-peak phase. For example, the turbulent kinetic energy is about $10\%$ larger in simulations with ${n_x=1}$ compared to those with ${n_x=4}$.

\subsection{Tilt angle}
\label{subsec:angle}

\begin{figure*}
\centering
	\includegraphics[width=0.9\columnwidth]{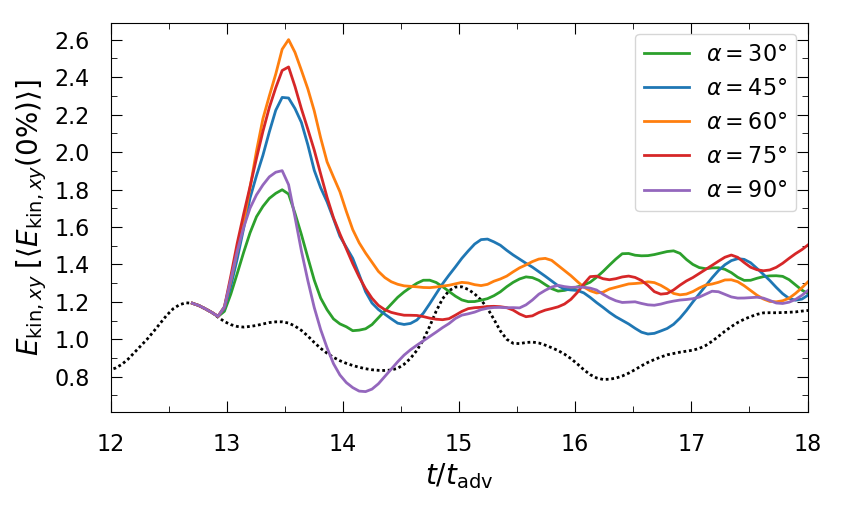}
	\includegraphics[width=0.9\columnwidth]{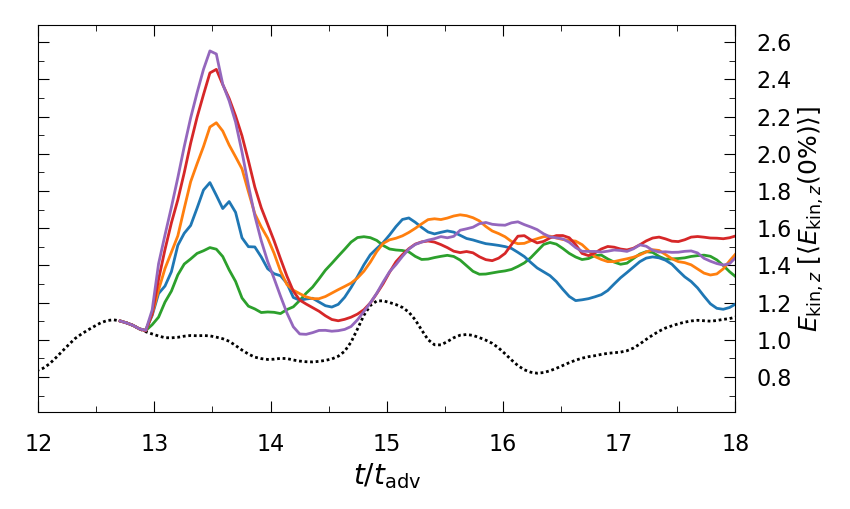}
	\caption{The horizontal (left-hand) and vertical (right-hand) components of the turbulent kinetic energy as a function of time for the simulations with varying tilt angle $\alpha$.
	See the models angX in Table~\ref{tab:simulations} for the rest of the parameters.
	All quantities are divided by the average values computed in the non-linear regime of neutrino-driven convection (black dotted curve). For readability, the data are smoothed out over a turnover time-scale.}
    \label{fig:Ekin_angle}
\end{figure*}

\begin{figure}
\centering
	\includegraphics[width=0.9\columnwidth]{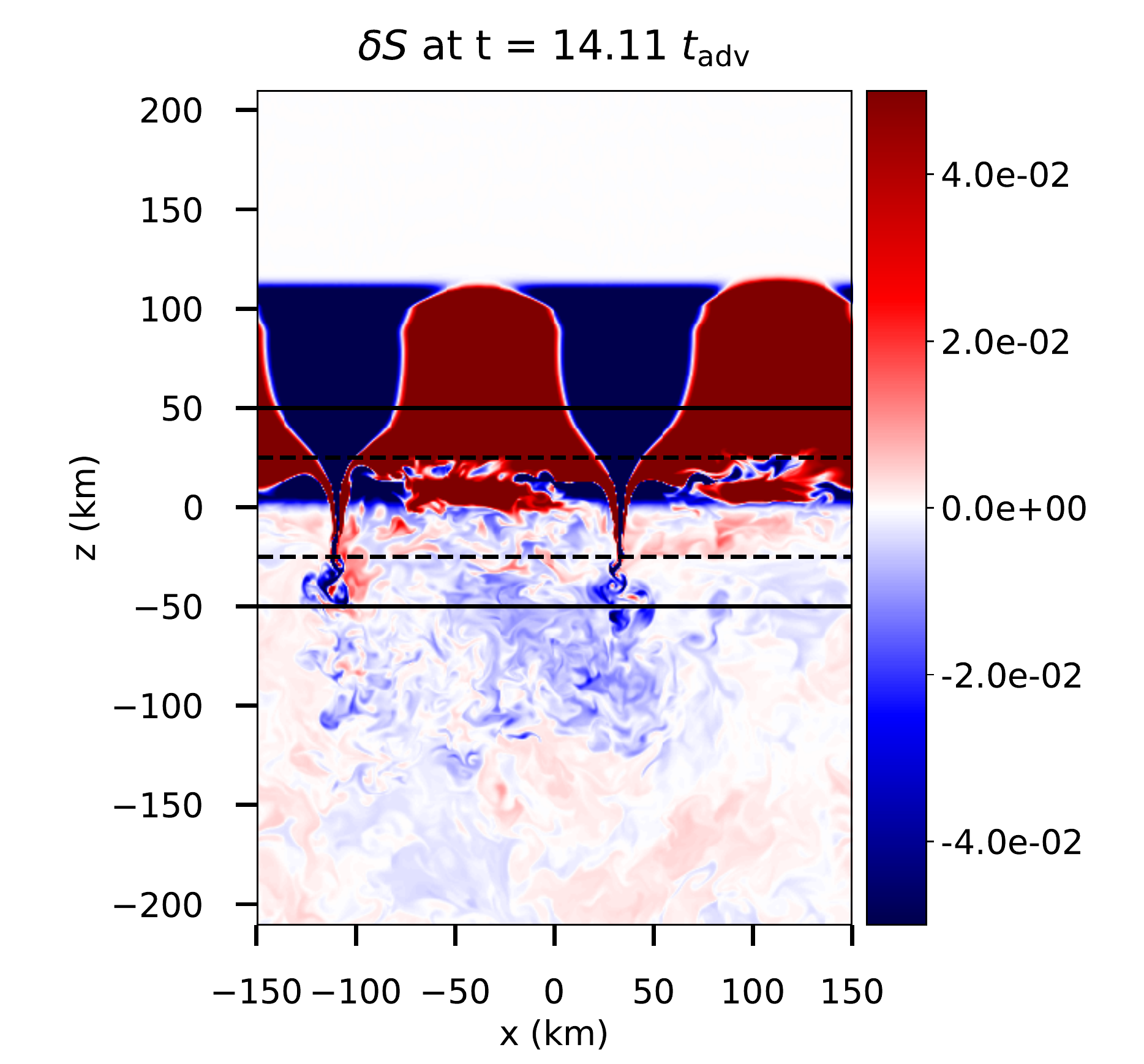}
    \caption{Snapshot in the plane ${y=0\,\rm{km}}$ of the simulation ang90 shown in entropy contrast ${\delta S=S-\langle S \rangle}$ and run with ${A=10\%}$, ${n_x=2}$, and ${\alpha=90\degree}$ (Table~\ref{tab:simulations}). At the time $t=14.11\,\tadv$, the injected perturbation has been partially accreted into the gain layer. The initially vertical entropy bands are deformed due to buoyancy force and adjustment to the new pressure. As a consequence, the bands of higher entropy material (in red) expand laterally and pinch the bands of lower entropy material (in blue) into narrow downflows reaching deeper regions of the gain layer than the rest of the perturbation. The horizontal lines are defined as in Fig.~\ref{fig:pert_sketch}.
    }
\label{fig:alpha90}
\end{figure}

\begin{figure}
\centering
	\includegraphics[width=0.9\columnwidth]{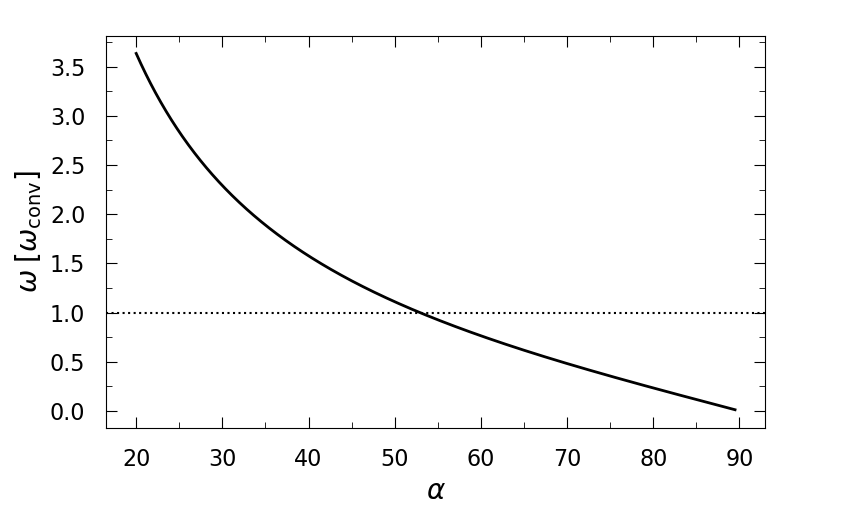}
    \caption{Temporal frequency $\omega$ of the perturbation, normalized by the convective turnover frequency $\omega_{\rm conv}$, as a function of the tilt angle $\alpha$ (solid curve).
    The horizontal dotted line depicts the equality between these two frequencies which is reached for ${\alpha_{\rm crit} \approx 53\degree}$.
    }
    \label{fig:omega_conv}
\end{figure}

In this section, we investigate the influence of the tilt angle $\alpha$ of the perturbation using the set of simulations with varying $\alpha$, labelled angX in Table~\ref{tab:simulations}. The other parameters are kept fixed: ${A=10\%}$ and ${n_x=2}$. The time evolution of $\Exy$ and $\Ez$ for these cases are shown in Fig.~\ref{fig:Ekin_angle}. 

The vertical component $\Ez$ increases with $\alpha$ (right-hand panel of Fig.~\ref{fig:Ekin_angle}). For example, the peak value of $\Ez$ for ${\alpha = 90\degree}$ is twice larger than that for ${\alpha=30\degree}$. This behavior can be explained by the structure of the perturbations. As $\alpha$ approaches $90\degree$, the entropy bands become more vertical and thus less prone to turbulent mixing. Instead, upon entering the gain region, the higher entropy regions inflate horizontally, generating narrow vertical downflows of colder material channeled to the bottom part of the gain layer, as shown in Fig.~\ref{fig:alpha90}. These downflows are sustained throughout the accretion of the perturbation, generating vertical motion.

In contrast, the turbulent kinetic energy in the horizontal directions does not grow monotonically with $\alpha$ (left-hand panel of Fig.~\ref{fig:Ekin_angle}). The maximum of $\Exy$ is reached for ${\alpha=60\degree}$ and is about $50\%$ higher than that reached for ${\alpha=90\degree}$. This behavior can be explained by comparing the temporal frequency $\omega$ of the perturbation, defined in Eq.~\ref{eq:omega}, with the convective turnover frequency $\omega_{\rm conv}$ in the gain layer. Following \citet{mueller16a}, we compute the average frequency $\omega_{\rm conv}$ during the non-linear phase of the convective instability as
\begin{equation}
 \label{eq:omega_conv}
 \omega_{\rm conv} = \dfrac{2\pi}{t_{\rm conv}} = \dfrac{2\pi v_{\rm conv}}{H} = \dfrac{2\pi}{H}\sqrt{\frac{2 \Ez}{M_{\rm g}}},
\end{equation}
where $v_{\rm conv}$ corresponds to the velocity of the convective turnovers. We use the non-perturbed model to calculate $\omega_{\rm conv}$. In this definition, we consider only the central ${-H/2 \leq z \leq H/2}$ part of the gain region, because the convective turnovers are mostly confined to this region. Fig.~\ref{fig:omega_conv} shows that perturbation frequency $\omega$ matches the convective turnover frequency $\omega_\mathrm{conv}$ at ${\alpha_{\rm crit} \approx 53\degree}$. This is close to the value ${\alpha=60\degree}$ that yields the largest $\Exy$ among our models (Fig.~\ref{fig:Ekin_angle}). This suggests that the coupling between the perturbation and the convection reaches a maximum efficiency for ${\omega \approx \omega_{\rm conv}}$. On the contrary, the peak of $\Exy$ decreases with increasing ${|\alpha - \alpha_{\rm crit}|}$. 

After the maximum of turbulent kinetic energy is reached, the horizontal and vertical components of the turbulent kinetic energy become roughly equal. There is no obvious correlation between $\alpha$ and the amount of turbulent kinetic energy at the late times.

\section{Analysis of the turbulent kinetic energy in the gain layer}
\label{sec:analysis}

\subsection{Qualitative model}
\label{subsec:origin}

Upon entering the gain region, the injected perturbations amplify the turbulent kinetic that region. In a quest to explain the mechanism for this amplification, \citet{mueller16a} considered several candidates, including the flux of acoustic energy ($F_{\rm ac}$), the conversion of gravitational potential energy into kinetic energy ($F_{\rm pot}$), and the flux of transverse kinetic energy ($F_{\rm t}$). Based on an order-of-magnitude estimate, they concluded that the additional turbulence generated by accreting perturbations mainly results from the work done by buoyancy. The simplicity of our setup allows us to verify this estimate. In our setup, $F_{\rm ac}$ is given by
\begin{equation}
 \label{eq:Fac}
 F_{\rm ac} = L_xL_y\delta P\delta v,
\end{equation}
where ${L_x=L_y=300\,\rm{km}}$ are the horizontal dimensions of the gain layer. $\delta P$ and $\delta v$ represent the pressure and velocity fluctuations with respect to horizontally averaged quantities.
The term $F_{\rm t}$ is computed as
\begin{equation}
 \label{eq:Ft}
 F_{\rm t} = \frac{1}{2} \dot{M} \delta v^2_{\rm h},
\end{equation}
where $\delta v_{\rm h}$ refers to the fluctuations of horizontal velocity. $\dot{M}$ corresponds to the mass accretion rate, defined as
\begin{equation}
    \label{eq:dM}
\dot{M}\equiv L_xL_y\rho|v_z|.
\end{equation}
Finally, $F_{\rm pot}$ is given by
\begin{equation}
 \label{eq:Fpot}
 F_{\rm pot} = \dot{M}\frac{\delta\rho}{\rho} \Delta\Phi.
\end{equation}
$\Delta\Phi$ is defined as the difference of gravitational potential between the upper and the lower edges of the gain layer.

\begin{figure}
\centering
	\includegraphics[width=0.9\columnwidth]{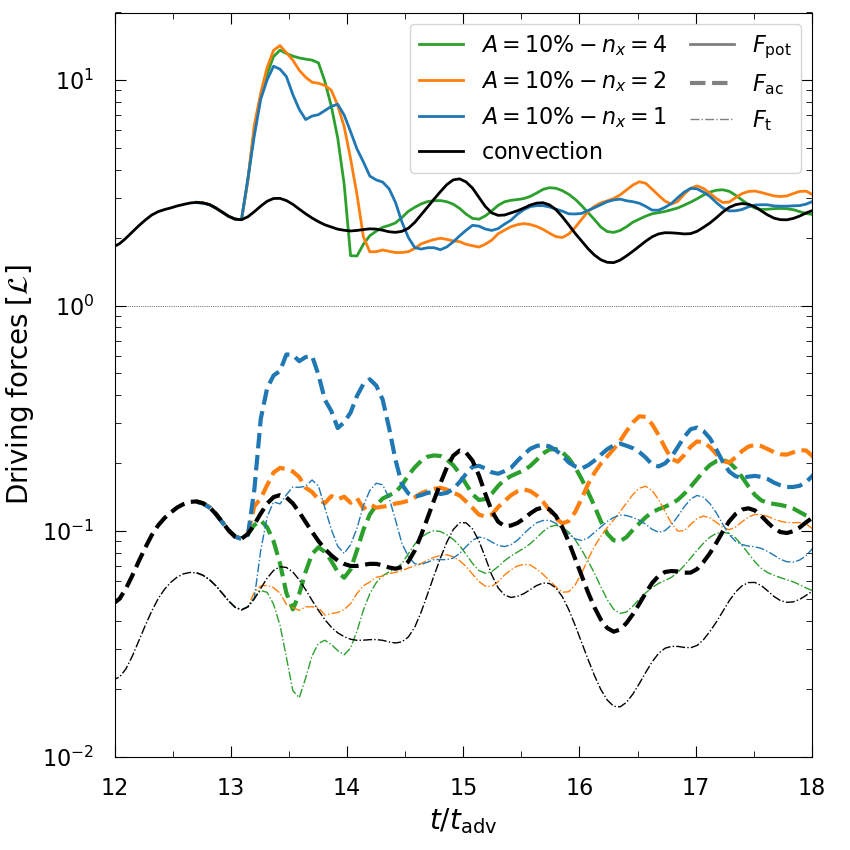}
    \caption{Time evolution of the driving terms $F_{\rm pot}$ (solid curves), $F_{\rm ac}$ (dashed curves), and $F_{\rm t}$ (dash-dotted curves) that could explain the origin of the extra turbulent kinetic energy during the perturbation accretion. These quantities are normalized by the total heating $\mathcal{L}$ which is depicted by the horizontal line. The simulation of convection (black) is compared to the simulations performed with $A=10\%$ and different horizontal wavenumbers: ${n_x=1}$ (blue), ${n_x=2}$ (orange), and ${n_x=4}$ (green).   
    }
    \label{fig:terms}
\end{figure}

We evaluate the three driving terms at the upper edge of the gain layer using our simulations performed with ${A=10\%}$ and with $n_x$ ranging from 1 to 4. Fig.~\ref{fig:terms} shows that $F_{\rm pot}$ exceeds $F_{\rm ac}$ and $F_{\rm t}$ by at least an order of magnitude, confirming the hypothesis of \citet{mueller16a}. Besides, $F_{\rm pot}$ becomes the dominant energy input over neutrino heating during the accretion of the perturbation.

In the absence of infalling perturbations, the turbulent kinetic energy in the gain layer can be obtained from Eqs.~\ref{eq:eta_conv_xy} and \ref{eq:eta_conv_z}: 
\begin{equation}
\label{eq:ekin_xy_nocorr}
\dfrac{\Exy}{M_{\rm g}} = \baretaconvxy \left[\dfrac{2H\mathcal{L}}{M_{\rm g}}\right]^{2/3}, 
\end{equation}
and 
\begin{equation}
\label{eq:ekin_z_nocorr}
\dfrac{\Ez}{M_{\rm g}} = \baretaconvz \left[\dfrac{2H\mathcal{L}}{M_{\rm g}}\right]^{2/3},
\end{equation}
where the average efficiency factors are ${\baretaconvxy \approx \baretaconvz \approx 0.3}$ in the non-linear phase of convection (cf. Fig.~\ref{fig:conv}). The increase of turbulent kinetic energy due to injection of perturbations can be accounted for via a correction term \citep{mueller16a, mueller17}:
\begin{equation}
\label{eq:ekin_xy_corr}
\dfrac{\Exy}{M_{\rm g}} = \baretaconvxy \left[\dfrac{2H\mathcal{L}}{M_{\rm g}}\right]^{2/3}\left(1+\psi\right)^{2/3}, 
\end{equation}
and 
\begin{equation}
\label{eq:ekin_z_corr}
\dfrac{\Ez}{M_{\rm g}} = \baretaconvz \left[\dfrac{2H\mathcal{L}}{M_{\rm g}}\right]^{2/3}\left(1+\psi\right)^{2/3},
\end{equation}
where
\begin{equation}
 \label{eq:psi}
 \psi \equiv \dfrac{\Lambda_{\rm p} F_{\rm pot}}{2H \mathcal{L}}.
\end{equation}
The parameter $\Lambda_{\rm p}$ measures the dissipation length of turbulence inside the injected perturbation. It corresponds to the driving scale of turbulent mixing which amounts to average vertical wavelength of the perturbation in the gain layer, i.e. $\lambda_z v_{\rm z,g}/v_{\rm up}$ (see Section \ref{subsec:duration}). This is thus different from the regime of neutrino-driven convection where the dissipation length is the height of the gain layer, i.e. $2H$. Note that $\Lambda_{\rm p}$ would correspond to the size of the infalling convective eddies in spherical geometry, i.e. ${\pi \rsh / \ell}$ \citep{mueller16a}.

Since the vertical wavelength of the perturbation grows with $\alpha$, it becomes larger than the gain layer height above a certain $\alpha$ (Eq.~\ref{eq:lambda_z}). However, the dissipation length cannot be larger than the gain region, so we limit the maximum value of $\Lambda_{\rm p}$ to $H$. This choice is justified for two reasons. First, \citet{mueller16a} advocated that the dissipation length associated to convection is limited to half of the gain layer height. Secondly, our simulations with large values of $\alpha$ show that turbulent mixing occurs only in the lower half of the gain layer, as it can be seen in Fig.~\ref{fig:alpha90}. The upper half is dominated by large bands of high-entropy material and narrow downflows without signs of turbulent mixing. Hence, we define the dissipation length as
\begin{equation}
 \label{eq:lambdap}
 \Lambda_{\rm p} \equiv \min{\left( \lambda_z \dfrac{v_{\rm z,g}}{v_{\rm up}}, H \right)} = \min{\left( \dfrac{L_x v_{\rm z,g}}{n_x v_{\rm up}}\tan{\left(\alpha\right)}, H \right)}.
\end{equation}
The threshold $\alpha$, above which ${\Lambda_\mathrm{p}=H}$, is
\begin{equation}
 \label{eq:alpha_thresh}
 \alpha_{\rm thresh} \equiv \arctan{\left(\dfrac{H n_x v_{\rm up}}{L_x v_{\rm z,g}}\right)}.
\end{equation}

\begin{figure*}
\centering
	\includegraphics[width=0.8\columnwidth]{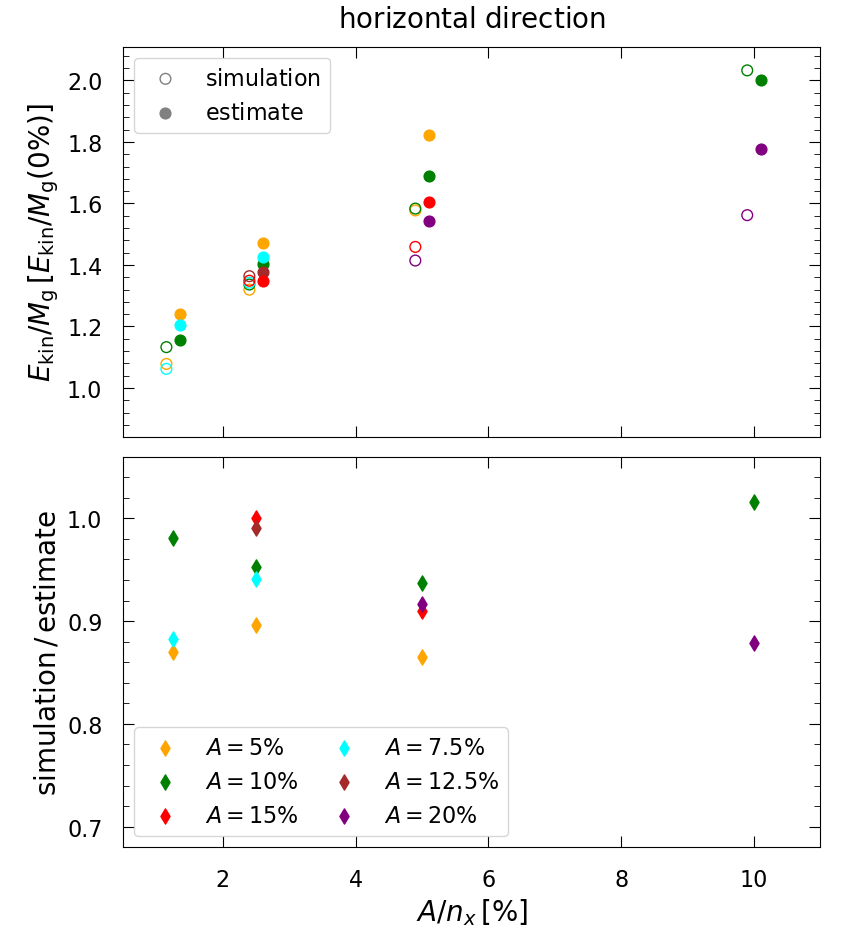}
	\includegraphics[width=0.8\columnwidth]{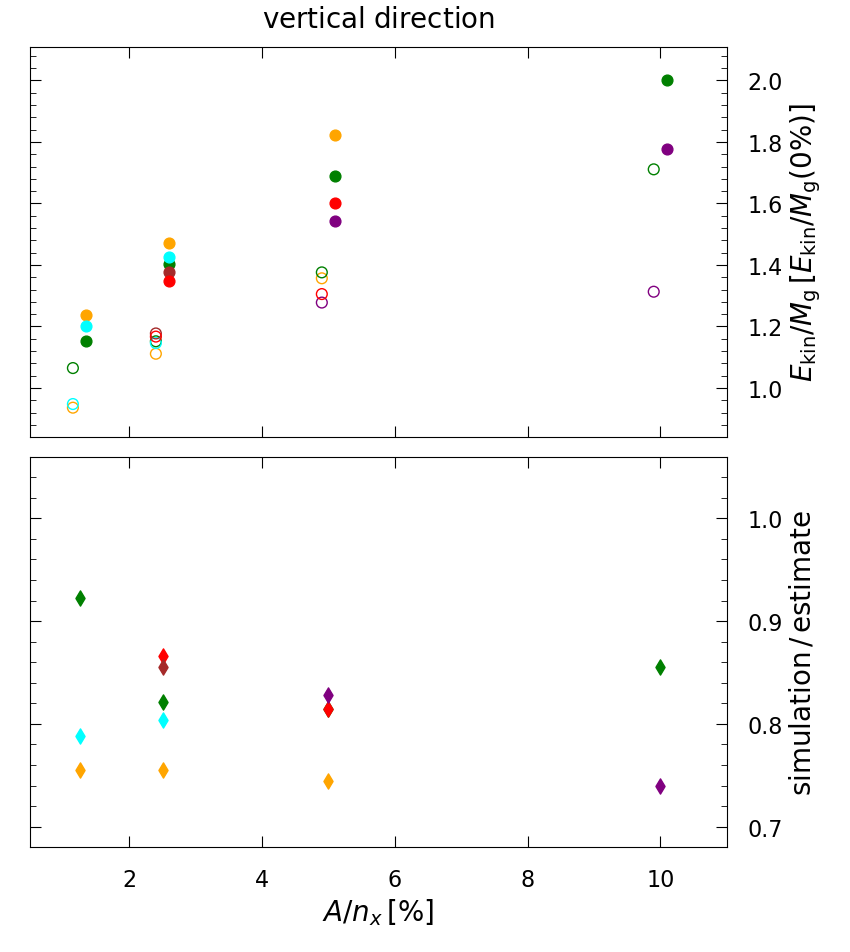}
	\caption{\textit{Upper panels}: The average ratios $\Exy/M_{\rm g}$ and $\Ez/M_{\rm g}$ measured in the simulations with varying $A$ and $n_x $ (models AX$\_$nxY in Table~\ref{tab:simulations}) during the perturbation accretion (open circles) are compared to the average estimates (filled circles) derived from Eqs.~\ref{eq:ekin_xy_corr} and \ref{eq:ekin_z_corr} for the horizontal direction (left-hand panel) and the vertical one (right-hand panel). The values of $\Exy/M_{\rm g}$ and $\Ez/M_{\rm g}$ shown here are normalized by the corresponding values of the non-perturbed model. These data are plotted as a function of the ratio $A/n_x$. Different perturbation amplitudes are represented by different colours (see lower legend). 
    \textit{Lower panels}: The ratios of average turbulent kinetic energy measured in the simulations to the estimates are shown as a function of the ratio $A/n_x$, respectively for the horizontal direction (left-hand panel) and the vertical one (right-hand panel).
	}
    \label{fig:comp_amplitude}
\end{figure*}

\begin{figure*}
\centering
	\includegraphics[width=0.8\columnwidth]{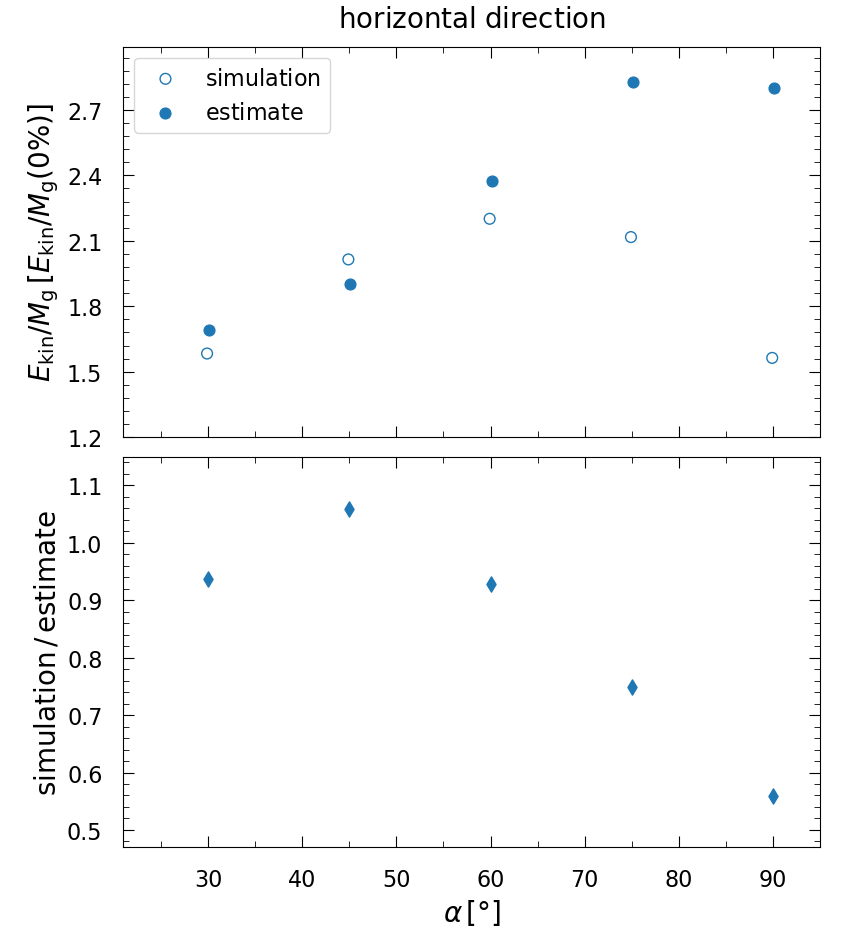}
	\includegraphics[width=0.8\columnwidth]{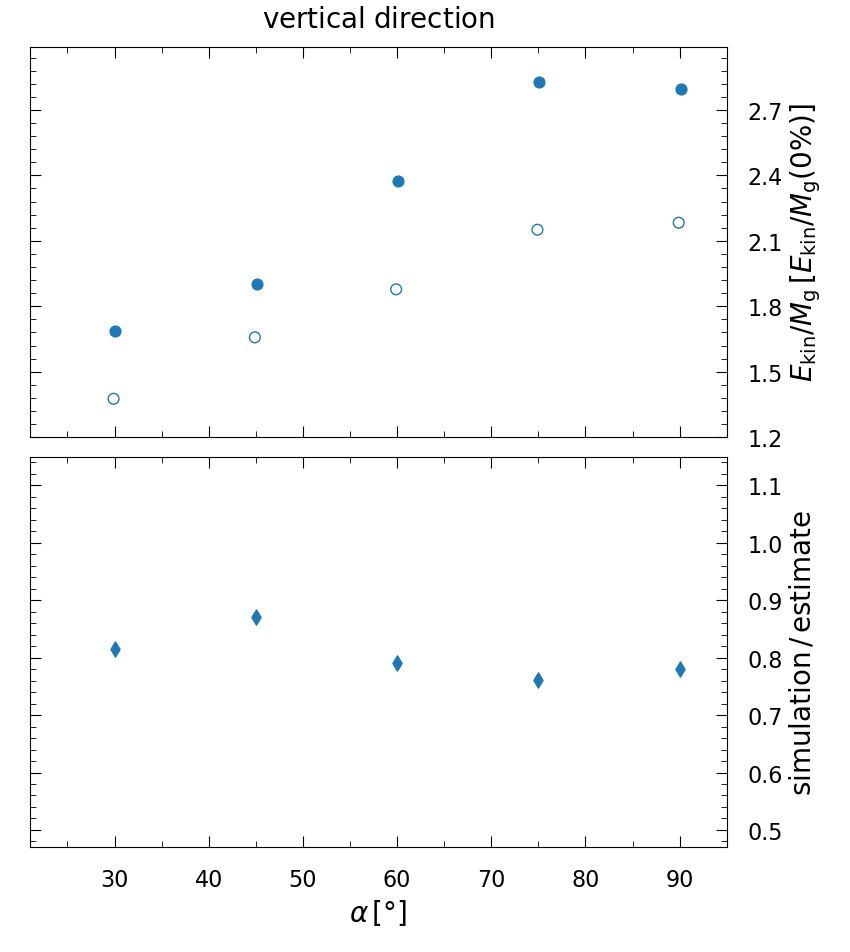}
	\caption{\textit{Upper panels}: The average ratios $\Exy/M_{\rm g}$ and $\Ez/M_{\rm g}$ measured in the simulations with varying $\alpha$ (models angX in Table~\ref{tab:simulations} during the perturbation accretion (open circles) are compared to the average estimates (filled circles) derived from Eqs.~\ref{eq:ekin_xy_corr} and \ref{eq:ekin_z_corr} for the horizontal direction (left-hand panel) and the vertical one (right-hand panel). The values of $\Exy/M_{\rm g}$ and $\Ez/M_{\rm g}$ shown here are normalized by the corresponding values of the non-perturbed model. These data are plotted as a function of the tilt angle $\alpha$. 
    \textit{Lower panels}: The ratios of average turbulent kinetic energy measured in the simulations to the estimates are shown as a function of the angle $\alpha$, respectively for the horizontal direction (left-hand panel) and the vertical one (right-hand panel).
	}
    \label{fig:comp_angle}
\end{figure*}

\subsection{Comparison to simulations}
\label{subsec:comparison}

In order to assess the robustness of the estimates introduced in the previous section, we compute the average values of the right-hand side of Eqs.~\ref{eq:ekin_xy_corr} and \ref{eq:ekin_z_corr} during the accretion phase of the perturbation, i.e. from ${t=13.2\,\tadv}$ to ${t=14.2\,\tadv}$, and compare these terms to the average quantities $\Exy/M_{\rm g}$ and $\Exy/M_{\rm g}$ measured in our simulations. In a more realistic setup, the additional turbulent pressure would result in wider gain region and larger neutrino heating, which cannot be reproduced in our stationary flow. Thus, we do not attempt to model the turbulent kinetic energy during the post-accretion phase of the perturbation.

The comparison for the set of models with varying $A$ and $n_x$, labelled AX$\_$nxY in Table~\ref{tab:simulations}, is displayed in Fig.~\ref{fig:comp_amplitude}. Note that ${\Lambda_{\rm p} < H}$ in all these models. We obtain agreement within ${\sim 10\%}$ and ${\sim 20\%}$ respectively for the horizontal (left-hand panels) and the vertical (right-hand panels) directions. In all but one case, the estimates are higher than the values measured in the simulations. A shorter and more elaborate dissipation length than $\Lambda_{\rm p}$ may improve the agreement even further.

Figure~\ref{fig:comp_amplitude} also confirms that the additional turbulent kinetic energy is mostly set by the ratio $A/n_x$. For a given ratio, the values measured in the corresponding simulations are close to each other. This result is particularly noticeable for the lowest ratios $A/n_x$ where different perturbations generate the same amount of turbulent kinetic energy within a few percents. Nevertheless, the discrepancies tend to widen at large ratio $A/n_x$, e.g. $A/n_x=10\%$, involving high amplitudes and spatial scales.
In these cases, large buoyant bubbles compress the region of active neutrino-driven convection which reduces the turbulent kinetic energy compared to simulations with smaller perturbation amplitudes (see Section~\ref{subsec:duration} for a description of this phenomenon). 

For most of the parameter space covered in our study, the turbulent kinetic energy increases by a factor of ${\sim 2}$ when an entropy perturbation accretes through the gain layer (Fig.~\ref{fig:comp_amplitude}). According to the Eqs.~\ref{eq:ekin_xy_corr} and \ref{eq:ekin_z_corr}, this increase corresponds to a factor ${\sim \left(1+2/3\psi\right)}$. \citet{mueller16b} proposed an order of magnitude estimate of $\psi$:
\begin{equation}
    \label{eq:psi_BM}
    \psi = \frac{\pi {\mathrm{Ma}}}{\ell \eta_\mathrm{acc}\eta_\mathrm{heat}},
\end{equation}
with $\eta_\mathrm{acc}$ being the accretion efficiency and $\eta_\mathrm{heat}$ the heating efficiency. In their 3D neutrino-hydrodynamics simulations, \citet{mueller17} measured ${\ell=2}$, ${\mathrm{Ma}=0.1}$, ${\eta_\mathrm{acc}=2}$, and ${\eta_\mathrm{heat}=0.05}$, which leads to a factor ${\left(1+ 2\psi/3\right)\sim 2}$, in agreement with our simulation results. The enhancement of the turbulent kinetic energy obtained in our simulations is thus in the ballpark of the results from more realistic models.

We now consider the sequence of models with varying $\alpha$, labelled angX in Table~\ref{tab:simulations}, for which ${\Lambda_{\rm p}=H}$ when ${\alpha \geq \alpha_{\rm thresh}\approx 68\degree}$. The results of the comparison are shown in Fig.~\ref{fig:comp_angle}. Overall, the agreement is similar to the previous comparison described above. They amount to about $10\%$ and $20\%$ respectively for the horizontal (left-hand panels) and the vertical (right-hand panels) directions. The agreement is less satisfactory for $\Exy$ at large angles. In these cases, the estimate becomes almost twice larger than the values measured in the simulations. 

In the horizontal direction (Fig.~\ref{fig:comp_angle}, left-hand panels), the turbulent kinetic energy is rather well approximated when ${\alpha < \alpha_{\rm thresh} \approx 68 \degree}$, that is when ${\Lambda_{\rm p} = \lambda_z v_{\rm up}/v_{\rm z,g}}$. The estimate is well suited to capture the trend of an increasing horizontal turbulent kinetic energy with the angle, up to the maximum located close to ${\alpha_{\rm crit} \approx 53 \degree}$. The gap between the simulations and the estimate becomes much wider when ${\alpha > \alpha_{\rm thresh}}$ and ${\Lambda_{\rm p}=H}$. The estimate results in a constant turbulent kinetic energy, independent from the angle, whereas $\Exy$ decreases from $60\degree$ to $90\degree$ in our simulations.
This decrease may be a consequence of a large $\omega$ that exceeds the convective turnover frequency. The temporal modulation of the perturbation, at large $\alpha$, may become too slow to affect neutrino-driven convection. This suggests that a more elaborate estimate could be designed by considering the distance between $\alpha$ and $\alpha_{\rm crit}$ in the definition of $\Lambda_{\rm p}$ (Eq.~\ref{eq:lambdap}). Our study shows that for the horizontal direction, the amount of turbulent kinetic energy is maximized when the convective turnover frequency matches $\omega$.

In the vertical direction, both the estimate of the turbulent kinetic energy and the values obtained in the simulations increase with larger $\alpha$. The estimate is about $20\%$ higher than the results of the simulations throughout the range of angles considered. $\Ez$ is an increasing function of $\alpha$ up to an asymptotic value reached for ${\alpha \sim \alpha_{\rm thresh}}$. In this regime of large angles, the size of the mixing region is not set by the vertical wavelength but is instead limited to half of the gain layer height. The maximum turbulent kinetic energy in the vertical direction is thus restricted by the size of the gain layer.

\subsection{Turbulent spectra}
\label{subsec:spectrum}

In order to characterize the change of spatial scales of the convective motions during the accretion phase of the perturbation, we examine the turbulent kinetic energy spectra of different simulations. The turbulent kinetic energy density is first decomposed into Fourier coefficients
\begin{equation}
    \label{eq:fourier}
    \hat{E}\left(k_x, k_y\right) = \int_{\Omega_{\rm h}} \exp{\left(-2\pi i \dfrac{k_x x + k_y y}{L_x}\right)} \sqrt{\rho}v_{\rm turb} \mathrm{d}\Omega,
\end{equation}
over each horizontal plane $\Omega_{\rm h}$, contained in the region ${-H \leq z \leq H}$ and then averaged over this height. $v_{\rm turb}$ is the turbulent velocity, defined as
\begin{equation}
    \label{eq:vturb}
    v_{\rm turb}^2 \equiv v_x^2 + v_y^2 + \left(v_z-\langle v_z \rangle \right)^2.
\end{equation}
To reduce the stochastic character of the turbulent spectra, we also average the Fourier coefficient over a time range ${\left[t,t+\tconv\right]}$.
For a given wavenumber $k$, the associated component of the power spectrum is then defined as
\begin{equation}
    \label{eq:power}
    E\left(k\right) = \sum_{k-1<||\left(k_x, k_y\right)||\leq k} |\hat{E}\left(k_x, k_y\right)|^2.
\end{equation}

\begin{figure}
\centering
	\includegraphics[width=0.9\columnwidth]{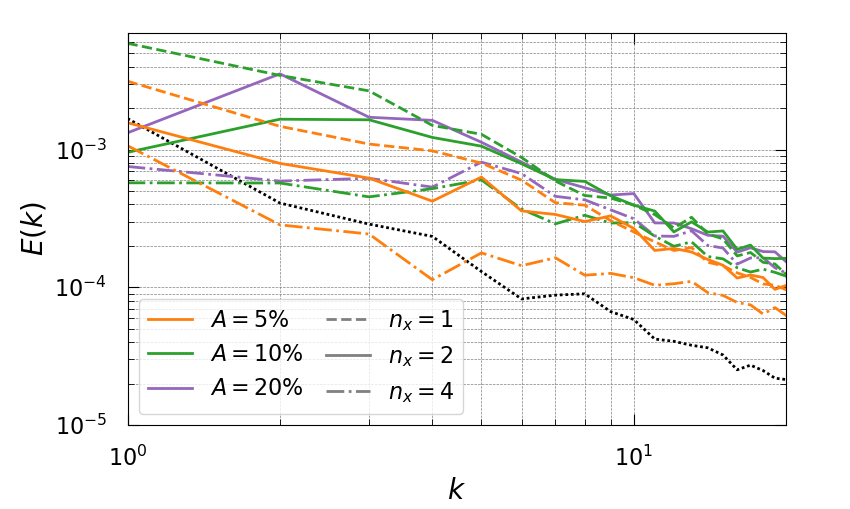}
    \caption{Spectra of turbulent kinetic energy computed in the models with varying $A$ and $n_x$ (models AX$\_$nxY in Table~\ref{tab:simulations}). The spectra are computed at the time when maximum turbulent kinetic energy is reached. The spectrum obtained in the simulation of non-perturbed convection is shown with a black dotted curve.}
    \label{fig:turb_amplitude_zoom}
\end{figure}

\begin{figure}
\centering
	\includegraphics[width=0.9\columnwidth]{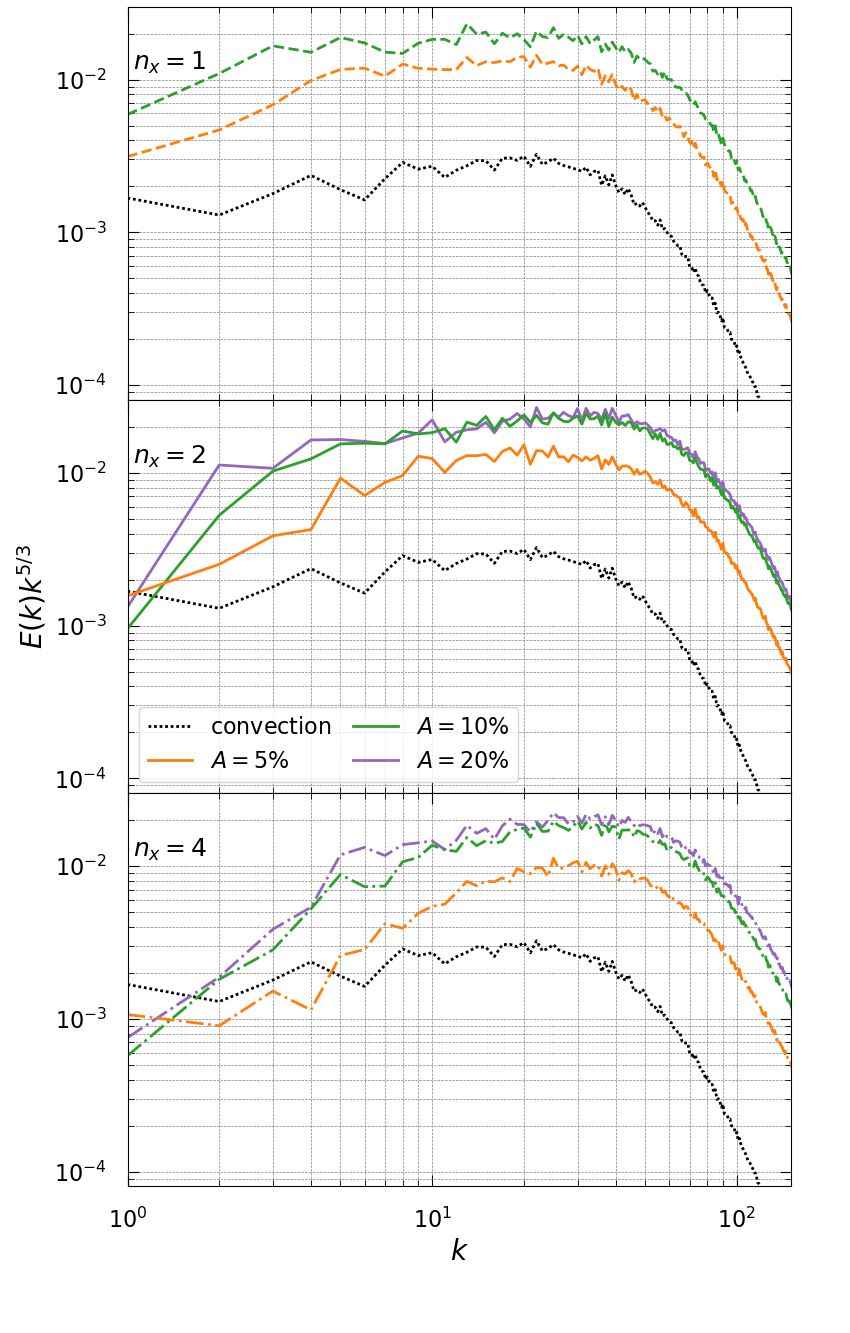}
    \caption{Spectra of turbulent kinetic energy compensated by $k^{5/3}$ for models with varying $A$ and $n_x$ (models AX$\_$nxY in Table~\ref{tab:simulations}). The simulations performed respectively with ${n_x=1}$, ${n_x=2}$, and ${n_x=4}$ are compared to the reference simulation (black dotted curve) in the upper, middle, and lower panels.
    }
    \label{fig:turb_amplitude_nx}
\end{figure}

First, we focus on the turbulent spectra for the models AX$\_$nxY that have varying $A$ and $n_x$ (cf. Table~\ref{tab:simulations}). Fig.~\ref{fig:turb_amplitude_zoom} shows the turbulent spectra taken at the time when the maximum turbulent kinetic energy is reached. The spectra of perturbed cases exceed that of the non-perturbed model. The only outlier is the simulation with ${A=5\%}$ and ${n_x=4}$, which exhibits a negligible increase of turbulent kinetic energy (Fig.~\ref{fig:Ekin_amplitude}). The scales at which the spectra $E(k)$ reach their peaks are closely related to the structure of the perturbations, at least when ${A\geq 10\%}$. The spectra of the models with ${n_x=2}$ show a peak at ${k=2}$, while the ones with ${n_x=4}$ display a peak at ${k=5}$. On the contrary, the spectra of all models with ${A=5\%}$ peak at ${k=1}$. These spectra resemble that of the non-perturbed convection which also peaks at ${k=1}$. This could be due to a faster disruption of such perturbations which rapidly leads to a system dominated by the spatial scales of the (unperturbed) neutrino-driven convection. 

Interestingly, spectra for models with a given ratio $A/n_x$ seem to overlap at intermediate scales, e.g. between ${k=2}$ and ${k=10}$. This is consistent with the observation that the turbulent kinetic energy is mostly set by the ratio $A/n_x$ (see Section~\ref{subsec:amplitude}). However, these overlaps are not systematic and not always clearly visible. For example, the case A10$\_$nx2 seems closer to A20$\_$nx2 instead of A20$\_$nx4. That said, these overlaps involve spatial scales that are subdominant in the turbulent kinetic energy budget. 

In Paper I, it was shown that the spectrum of unperturbed 3D neutrino-driven turbulent convection scales as  $\propto k^{-5/3}$ for ${10\lesssim k \lesssim 30}$ \citep[see also][]{radice16, nagakura19, melson20}. In order to identify the inertial range of turbulence in perturbed models, we show the compensated spectra $E(k) k^{5/3}$ in Fig.~\ref{fig:turb_amplitude_nx}. We find that an inertial range, where the function $E(k) k^{5/3}$ is flat, can clearly be observed only for ${n_x=1}$. For larger $n_x$, the inertial range is smaller. The perturbations with ${n_x=1}$ reproduce more closely the properties of the turbulence generated by (unperturbed) neutrino-driven convection, which favours the largest spatial scales \citep{kazeroni18}. 

\begin{figure}
\centering
	\includegraphics[width=0.9\columnwidth]{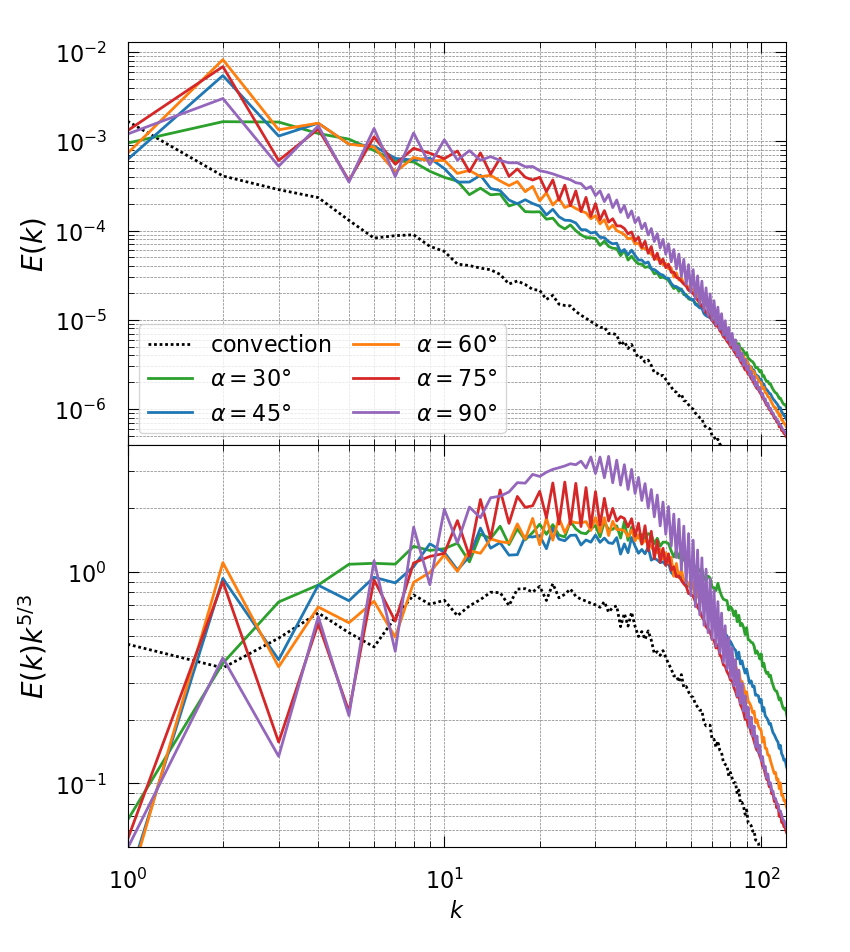}
    \caption{\textit{Upper panel}: Spectra of turbulent kinetic energy for models with varying $\alpha$ (models angX in Table~\ref{tab:simulations}). The spectra are computed at the time when maximum turbulent kinetic energy is reached. The spectrum obtained in the simulation of non-perturbed convection is shown with a black dotted curve.
    \textit{Lower panel}: Same as above but for spectra compensated by $k^{5/3}$.
    }
    \label{fig:turb_angle}
\end{figure}

Figure~\ref{fig:turb_angle} shows the turbulent spectra of models with varying $\alpha$, labelled angX in Table~\ref{tab:simulations}. All perturbed cases display a peak in their spectra at ${k=2}$, which corresponds to the input parameter ${n_x=2}$ (top panel). The peak is maximized for ${\alpha=60\degree}$, which is consistent with the angle-dependence of the turbulent kinetic energy that we observed earlier in Fig.~\ref{fig:Ekin_angle}. The comparison of these spectra shows that the total turbulent kinetic energy is again maximized for angles such that $\omega \approx \omega_{\rm conv}$. 

The spectra of the models with large $\alpha$ are characterized by a pronounced "odd-even" deformation. As the angle approaches $90\degree$, the bands of entropy become vertical and this prevents efficient disruption of the perturbation, as it was illustrated in Fig.~\ref{fig:alpha90}. Instead, the perturbation is dominated by the initial spatial scale ${n_x=2}$ throughout its accretion into the gain layer. We hypothesize that this behavior is the reason for the "odd-even" shape of the spectra. The large angles not only enhance the modulations of the spectra, but also reduce the range of wavenumbers for which an inertial range could be identified (cf. the bottom panel of Fig.~\ref{fig:turb_angle}). Such a range is hardly noticeable for ${\alpha \geq 60\degree}$.

\section{Impact of the dimensionality}
\label{sec:dimension}

\begin{figure*}
\centering
	\includegraphics[width=0.9\columnwidth]{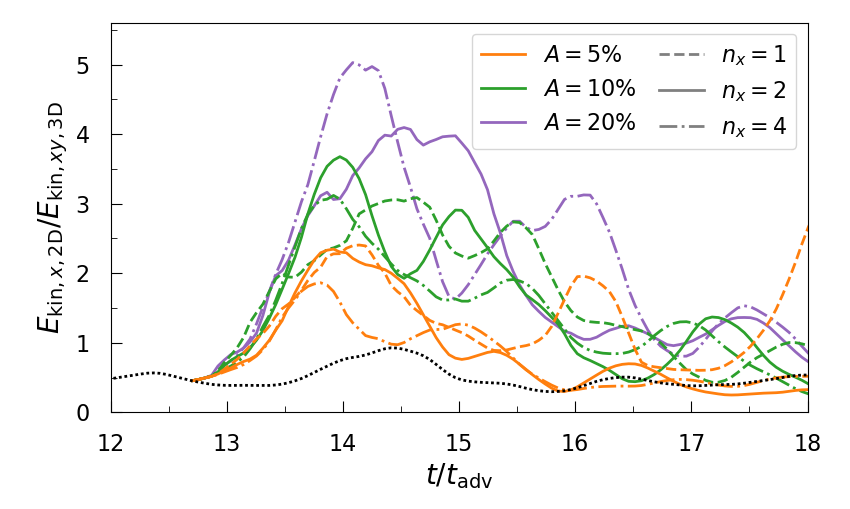}
	\includegraphics[width=0.9\columnwidth]{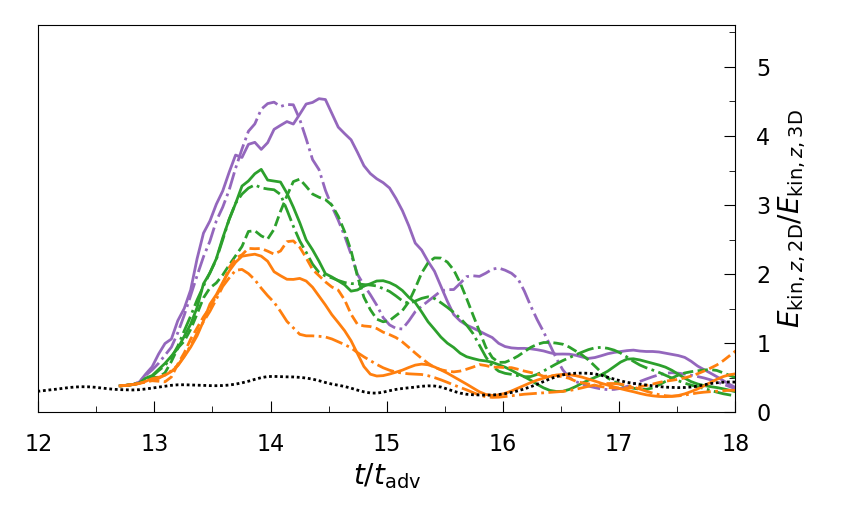}
	\caption{
	Ratio of the horizontal (left-hand panel) and the vertical (right-hand panel) components of the turbulent kinetic energy between 2D and 3D simulations for the sequence with varying $A$ and $n_x$ as a function of time. See the models AX$\_$nxY in Table~\ref{tab:simulations} for the rest of the parameters. 
    For readability, the data are smoothed out over a turnover time-scale.}
    \label{fig:Ekin_2D3D}
\end{figure*}

\begin{figure*}
\centering
	\includegraphics[width=0.8\columnwidth]{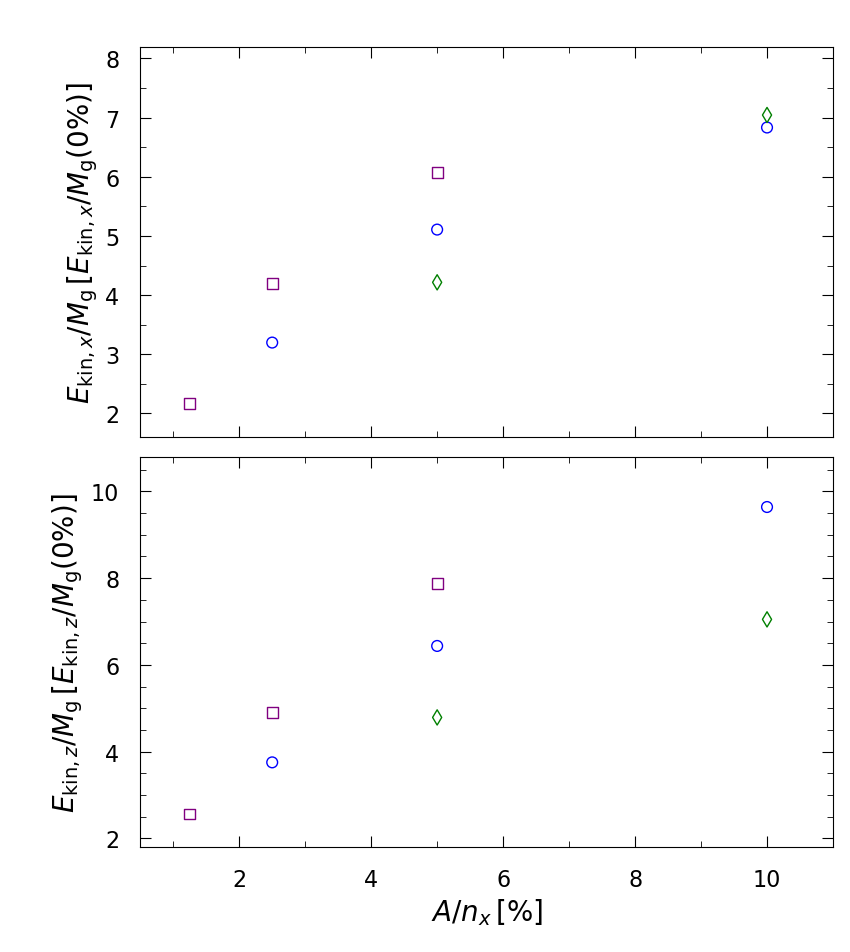}
	\includegraphics[width=0.8\columnwidth]{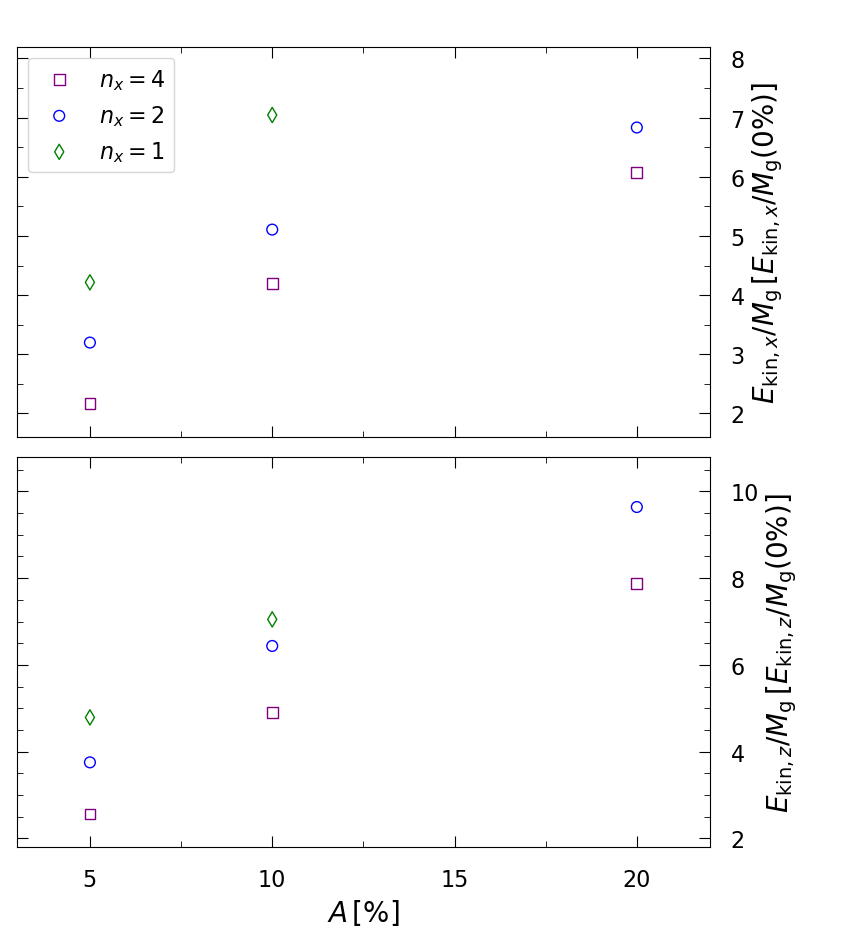}
	\caption{The average ratios $E_{\mathrm{kin},x}/M_{\mathrm{gain}}$ (upper panels) and $E_{\mathrm{kin},z}/M_{\mathrm{gain}}$ (lower panels) measured in the 2D simulations with varying $A$ and $n_x$ are plotted against the ratio $A/n_x$ (left-hand panels) and the perturbation amplitude $A$ (right-hand panels). The values of $E_{\mathrm{kin},x}/M_{\rm g}$ and $\Ez/M_{\rm g}$ shown here are normalized by the corresponding values of the non-perturbed model. See the models AX$\_$nxY in Table~\ref{tab:simulations} for the rest of the parameters. 
	}
    \label{fig:comp_Ekin_2D}
\end{figure*}

In order to study the influence of the dimensionality on our results, we simulate in 2D both the unperturbed model and the sequence of models with varying $A$ and $n_x$ (labelled AX$\_$nxY in Table~\ref{tab:simulations}). Fig.~\ref{fig:Ekin_2D3D} shows the ratio of the components of the turbulent kinetic energy between 2D and 3D. For the non-perturbed model, the turbulent kinetic energy is twice larger in 3D than in 2D. This is different from the results of Paper I where a stationary flow with ${\chi=5}$ was considered and the turbulent kinetic energy was found to be slightly larger in 2D. In our current work, we consider a flow with ${\chi=3}$ which is closer to the marginal stability. For such a marginally unstable flow, convection is triggered more rapidly in 3D and buoyant bubbles reach higher altitudes and velocities \citep{kazeroni18}. 

The 2D nature of the flow has a profound impact on the turbulence generated by an accreting perturbation. The turbulent kinetic energy measured in 2D simulations is between $2$ to $5$ times larger than that in 3D runs, as shown in Fig.~\ref{fig:Ekin_2D3D}. The gap between 2D and 3D increases the perturbation amplitude. On the other hand, the horizontal wavenumber does not significantly affect the gap. In 2D, the disruption of the injected perturbation leads to the formation of large scale vortices that tend to merge until reaching the gain layer height. Once formed, such a flow pattern may remain in the gain layer for several advection times before being expelled in the downstream region \citep{kazeroni18}. For this reason, large peaks of turbulent kinetic energy can be observed long after the perturbation has fully accreted into the gain region (Fig.~\ref{fig:Ekin_2D3D}). 

Contrary to the 3D simulations, the turbulent kinetic energy in 2D simulations does not correlate with the ratio $A/n_x$ (Fig.~\ref{fig:comp_Ekin_2D}, left-hand panels). For a given $A/n_x$, the mean values of $\Ex/M_{\rm g}$ and $\Ez/M_{\rm g}$ are spread over a wide range that overlaps with the distribution of values obtained for other values of $A/n_x$. Both $\Ex/M_{\rm g}$ and $\Ez/M_{\rm g}$ increase with lowering $n_x$ (Fig.~\ref{fig:comp_Ekin_2D}, right-hand panels). At the same time, there is no clear correlation between the perturbation amplitude $A$ and the ratios $\Ex/M_{\rm g}$ and $\Ez/M_{\rm g}$. This is related to the inverse turbulent cascade that takes place in 2D. When a perturbation with ${n_x=1}$ accretes into the gain region, its geometry already corresponds to that of the largest possible scale of the convection. This results in a more rapid growth of the turbulence across the gain region. For this reason, the amplitude is less important than $n_x$.

\section{Conclusions and Discussion}
\label{sec:conclusion}

We have studied the coupling between injected density perturbations and neutrino-driven convection in the gain layer of a CCSN, employing parametrized 3D simulations of an idealized model. Our study considered a stationary flow where neutrino-driven convection has reached the non-linear stage \citep{kazeroni18} and into which density perturbations are injected. The latter originates from the interaction between pre-collapse inhomogeneities and the stalled shock of a CCSN. The amplitude, the dominant spatial scale, and the temporal frequency of the injected density perturbations were varied in a systematic manner to characterize the additional turbulence generated by the accretion of these fluctuations. A comparison between 2D and 3D simulations was conducted to clarify the impact of the dimensionality on the additional turbulence. Our main findings can be summarized as follows. 
\newline

1. The turbulent kinetic energy in the gain layer scales as $A/n_x$, where $A$ is perturbation amplitude and $n_x$ is the horizontal wavenumber. For example, doubling $A$ or halving $n_x$ result in a similar increase of the turbulent kinetic energy. As a consequence, the impact of small-scale perturbations is weaker. This supports the hypothesis that only asymmetries at the largest spatial scales, e.g. ${\ell=1-2}$, could be beneficial to revive the shock wave \citep[e.g.][]{mueller15a, mueller16a}.
\newline

2. The influence of the temporal frequency of the perturbation on neutrino-driven convection was investigated by varying the tilt angle $\alpha$ between the horizontal axis and the direction of the entropy bands of the perturbation. In the horizontal direction, the turbulent kinetic energy is maximized when the frequency of the perturbation is close to that of the convective turnovers. The vertical component increases monotonically with $\alpha$. When ${\alpha\approx 90\degree}$, the vertical entropy bands become less prone to turbulent mixing and their disruption leads to greater vertical displacements of matter which increases the turbulent kinetic energy in that direction. Our work suggests that density perturbations may non-linearly increase the turbulent kinetic energy of the convective instability by a few tens of percents for a wide range of perturbation frequencies. 
\newline

3. We assessed the accuracy of the qualitative model of the turbulent kinetic energy generated by an accreting perturbation by \citet{mueller16a, mueller17}. An agreement within $20\%$ was obtained between the simulations and the estimate for most of our parameter space. The only exception arises at large tilt angles, for which low perturbation frequencies barely affect neutrino-driven convection. Our study unambiguously demonstrates that the additional turbulent kinetic energy originates from the work of buoyancy onto the density perturbations, at least when the neutrino-driven convection dominates the post-shock dynamics, in agreement with \citet{mueller16a, mueller17}.
\newline

4. The turbulent kinetic energy in the gain layer increases with the scale of the perturbation both in 2D and in 3D. However, 2D simulations overestimate the additional turbulent kinetic energy by a factor of ${\sim 2-5}$ compared to their 3D counterparts. This difference stems from the direction of the turbulent energy cascade \citep{murphy13, kazeroni18}. Unlike 3D models, the fluctuations are not disrupted into smaller scales in 2D. This leads to a rapid growth of large-scale turbulent motions, resulting in unphysically large turbulent kinetic energies (cf. Section~\ref{sec:dimension} for more details).
\newline

Our study represents an important step towards a comprehensive description of the interaction of pre-collapse fluctuations with non-linear instabilities in the gain layer of CCSNe. We have clarified the origin and the properties of the additional turbulence related to the coupling between density perturbations and neutrino-driven convection. Our results are in line with the few 3D CCSN simulations performed from the final minutes of the progenitor before collapse to the shock revival. Simplified models remain essential to analyze in depth the turbulence resulting from the interplay between different types of pre-collapse fluctuations and the variety of post-shock instabilities while covering a broad parameter space.

In our study, we neglected the role played by the stationary shock. The inclusion of the latter could lead to a subtle interplay between SASI and neutrino-driven convection as witnessed by \citet{mueller17}. In that study, the accreting perturbation, coupled to the shock oscillations, excited turbulent motions primarily in the radial direction whereas turbulence was enhanced equivalently in all directions for our neutrino-driven convection dominated flow. Adding a shock wave to our model could reveal how neutrino-driven convection and SASI are enhanced or damped by an accreting perturbation. Our setup including a stationary shock would also be insightful to quantify the reduction of the critical luminosity for any type of accreting perturbation interacting with a given post-shock instability. This will be the subject of the future studies.

\section*{Acknowledgments}

The authors thank Thierry Foglizzo for productive discussions. EA acknowledges support from the Ministry of Education and Science of the Republic of Kazakhstan through the grants BR05236454 and AP05135753. Computer resources for this project have been provided by the Max Planck
Computing and Data Facility (MPCDF) on the HPC system Cobra.


\bibliographystyle{mnras}
\bibliography{convection_bibli} 



\appendix


\bsp	
\label{lastpage}
\end{document}